\documentclass[a4paper,12pt,english]{article}

\usepackage{amsfonts,bm,amssymb,euscript,array,babel,cite}

\usepackage{amsmath,amsthm}
\usepackage[dvips]{graphicx,color}
\usepackage[T1]{fontenc}
\usepackage{framed}

\usepackage{empheq}

\makeatletter

\newlength\dlf  

\newcommand*{\dittostraight}{---\textquotedbl---}

\newcommand{\dd}{\mathrm{d}}
\newcommand{\e}{\mathrm{e}}
\newcommand{\w}{\wedge}
\newcommand{\bbm}{\left(\begin{matrix}}
\newcommand{\ebm}{\end{matrix}\right)}
\newcommand{\beq}{\begin{eqnarray}}
\newcommand{\eeq}{\end{eqnarray}}

\makeatother

\def\mx{\mathfrak{X}}

\newtheorem{prop}{Proposition}[section]

\newtheorem{defn}[prop]{Definition}

\newcommand{\sfrac}[2]{{\textstyle\frac{#1}{#2}}}

\newcommand{\be}{\begin{equation}}
\newcommand{\ee}{\end{equation}}

\newcommand{\beqa}{\begin{eqnarray}}
\newcommand{\eeqa}{\end{eqnarray}} 
\def\nn{\nonumber} \def \bea{\begin{eqnarray}} \def\eea{\end{eqnarray}}

\newcommand{\barr}{\begin{array}}
\newcommand{\earr}{\end{array}}
\numberwithin{equation}{section}
\newcommand{\mf}{\mathfrak}
\def\tb{\text{TM}}
\def\ctb{\text{T}^{\star}\text{M}}

  \def\G{\Gamma}
 \def\d{\delta} 

   \def\m{\mu}
\def\n{\nu} \def\o{\omega}   \def\r{\rho}
\def\s{\sigma} \def\S{\Sigma} \def\t{\tau} 
  \def\X{\Xeta}

   \def\X{\mathbb X}

 \def\one{\mbox{1 \kern-.59em {\rm l}}}

\def\bit{\begin{itemize}} \def\eit{\end{itemize}}

\def\({\left(} \def\){\right)}

 \allowdisplaybreaks[3]

\begin{document}

\makeatother


\parindent=0cm

\renewcommand{\title}[1]{\vspace{10mm}\noindent{\Large{\bf

#1}}\vspace{8mm}} \newcommand{\authors}[1]{\noindent{\large

#1}\vspace{5mm}} \newcommand{\address}[1]{{\itshape #1\vspace{2mm}}}


\begin{titlepage}

\begin{flushright}
\today \\
ITP-UH-07/15
\end{flushright}

\begin{center}

\vskip 3mm

\title{ {\Large 
Sigma models for genuinely non-geometric backgrounds}}

\vskip 3mm

 \authors{\large 
 Athanasios {Chatzistavrakidis}$^{\sharp,}\footnote{thanasis@itp.uni-hannover.de}$,  Larisa Jonke$^{\sharp,\dagger,}\footnote{larisa@irb.hr}$, Olaf Lechtenfeld$^{\sharp,}\footnote{lechtenf@itp.uni-hannover.de}$}

\vskip 3mm

\address{ {$^{\sharp}$}\
Institut f\"ur Theoretische Physik,  
Leibniz Universit\"at Hannover,\\  Appelstra{\ss}e 2, 30167 Hannover, Germany 

\

{ {$^{\dagger}$}\
 Division of  Theoretical Physics, 
 Rudjer Bo$\check s$kovi\'c Institute, \\
 Bijeni$\check c$ka 54, 10000  Zagreb, Croatia}
}

\bigskip 

\end{center}

\vskip 2cm

 \begin{center}
\textbf{Abstract}

\vskip 3mm

 \begin{minipage}{14cm}%

The existence of genuinely non-geometric backgrounds, i.e. ones without geometric dual, is an important question in string theory. 
In this paper we 
examine this question from a sigma model perspective. First we construct a particular class of Courant algebroids as 
protobialgebroids with all types of geometric and non-geometric fluxes. For such structures we apply the mathematical result that any 
Courant algebroid gives rise to a 3D topological sigma model of the AKSZ type and we discuss the corresponding 2D field theories. 
It is found that these models are always geometric, even when both 2-form and 2-vector fields are neither vanishing nor inverse 
of one another. Taking a further step, we 
suggest an extended class of 3D sigma models, whose world volume is embedded in phase space, which allow for 
genuinely non-geometric backgrounds. Adopting the doubled 
formalism such models can be related to double field theory, albeit from a world sheet perspective.

 \end{minipage}
 \end{center}

\end{titlepage}

\tableofcontents

\section{Introduction} 

It is by now well established that useful string vacua, in particular those where potentials that can stabilize the moduli are generated, 
contain background fluxes. These fluxes come in several types, in particular standard ones such as NSNS flux, 
torsion or geometric flux and RR fluxes, but also non-standard types such as 
non-geometric fluxes in the NSNS \cite{Shelton:2005cf} and RR \cite{Aldazabal:2006up} sectors. 
The latter can be described with techniques from the differential 
geometry of Lie and Courant algebroids
\cite{Halmagyi:2008dr,Halmagyi:2009te,Blumenhagen:2012pc,Mylonas:2012pg,Blumenhagen:2013aia,Chatzistavrakidis:2013wra}
which are also used in Hitchin's generalized complex geometry \cite{gcg2}. They often appear in 
(generalized) T- or S-duals of standard geometries \cite{Hull:2004in,Hull:2007jy,Hull:2009sg,Dabholkar:2005ve}. 
However, in most studies up to now these non-geometric string backgrounds 
are not truly new string vacua, even referring to ideal cases where the string equations of motion are/could be solved. 
This is so because the very essence of dualities is that the physics is the same on each side, 
and thus vacua which are related by dualities are just different ways to describe the same physics.
This can change only by having at hand vacua which are genuinely non-geometric, 
which means there is no duality transformation that can map the vacuum to a known, geometric one. 
In this sense, cases known as the Q-background and the R-background are not truly non-geometric from a string-theoretic 
point of view.

The main question that we would like to address in this paper
is how genuinely non-geometric models could be described.
 This is arguably the most essential question in the study of generalized flux compactifications, which however has 
been posed and addressed only fragmentarily. One direction which was followed relies on the construction of exact conformal field 
theories based on asymmetric orbifolds \cite{Narain:1986qm}. Asymmetric orbifolds are backgrounds where left- and right-moving 
string coordinates see different geometries; for this reason such theories are indeed genuinely non-geometric 
string solutions and they can contain all types of fluxes \cite{Hellerman:2002ax,Hellerman:2006tx,Schulgin:2008fv,Condeescu:2012sp,Condeescu:2013yma}.
Other approaches on the same problem include work on heterotic string vacua with non-geometric fluxes \cite{McOrist:2010jw}, 
where it is argued that geometric and non-geometric compactifications are equally typical, the study of cases where not only the 
internal geometry but also the external one is multi-valued and thus non-geometric \cite{deBoer:2012ma}, as well as a 
classification of the U-duality orbits of gaugings of (half-)maximal supergravities \cite{Dibitetto:2012rk}.

Our approach is different than the above ones and complements previous work in the string theory literature 
\cite{Halmagyi:2008dr,Halmagyi:2009te,Blumenhagen:2012pc,Mylonas:2012pg,Blumenhagen:2013aia,Chatzistavrakidis:2013wra}.
The starting point is Courant algebroids (CAs), which are structures introduced in Ref. \cite{wein} that 
provide a systematization of the properties of the Courant bracket introduced in Ref. \cite{dirac}. 
The authors of \cite{wein} construct CAs as Lie bialgebroids, which is a special case of a more general 
construction performed in Ref. \cite{Roy} using the notion of protobialgebroids (PBAs). The latter are structures that incorporate 3-index twists, corresponding to (some of) the NSNS fluxes that appear in string theory. In this paper we construct a class of PBAs, putting on firmer grounds and generalizing our previous
work \cite{Chatzistavrakidis:2013wra}. We consider PBAs following the spirit of twisting the 
generalized tangent bundle $\tb\oplus\ctb$ of a $d$-torus by 2-form, 2-vector and (1,1)-tensor 
deformations. We choose a representative paradigm of (1,1) deformations that leads to nilmanifolds, which 
are also termed ``twisted tori'' in physics; this is a natural way to go beyond the toroidal case. This approach directly suggests how brackets, morphisms and generalized 3-forms 
should be defined in the class of PBAs we study. Given that any PBA gives rise to a CA, we then proceed in 
the construction of the latter and discuss an illustrative example of the class. The twist 
approach that we follow yields non-standard CAs.

Having constructed the desired structures over twisted tori, it is desirable for physics to study sigma models 
that correspond to them. An important mathematical result (\cite{Roytenberg:2006qz}) states that given a CA one can construct a 
topological sigma model of the type introduced in the seminal work of Alexandrov, Kontsevich, Schwarz and 
Zaboronsky (AKSZ) \cite{Alexandrov:1995kv} (see Ref. \cite{ikeda} for a useful review). Applying this result, 
we construct in very explicit terms the sigma model corresponding to the class of CAs with all types of twists. 
Considering that the 3D topological theory has a 2D boundary, we derive consistency conditions for the 
action including all twists and deformations, thus generalizing previous results. Moreover, we 
show that in certain limits these conditions reproduce known results, for example integrability conditions 
for Dirac structures, such as the ones found in Refs. \cite{severa,sw,Hofman:2002rv} and additional 
ones derived in Ref. \cite{Chatzistavrakidis:2013wra}. Additionally, we discuss the corresponding 
2D theories, add dynamics, and discuss in explicit detail a particular example with all types of fluxes.

After presenting models with all kinds of twists and 2-form and 2-vector deformations that are neither 
vanishing nor inverse of one another, we discuss whether such cases can be considered non-geometric 
in the sense of string theory. In this discussion one has to invoke T-duality in order to differentiate 
between two different kinds of 3-vector $R$ flux, the one obtained by standard differential geometric 
methods (being a derivation of a non-Poisson 2-vector $\beta$, or equivalently the Schouten bracket of 
$\beta$ with itself) and the one obtained by generalized T-duality. As already noticed in Ref.
 \cite{Halmagyi:2009te} the former does not deserve to be called non-geometric, since it can be 
 associated to a 2D sigma model with a standard target space. However, it is the second kind of 
 $R$-flux that appears in string theory after T-duality. This means that a generalization of the 
 formalism we apply in sections 2 and 3 is necessary in order to account for the $R$-flux originating in T-duality, and 
 in order to examine the possibility of genuinely non-geometric backgrounds.
 
 The above discussion leads us to propose an extension of the sigma models with standard target space 
 to ones with phase space as target. This is also motivated by a similar approach adopted in Ref. 
 \cite{Mylonas:2012pg}. We choose to work with reference to the doubled formalism of string theory 
 \cite{Tseytlin:1990nb}, which parallels the first order formalism on phase space and 
 introduces a set of coordinates $\tilde X_a$ in addition to the standard coordinates $X^a$, 
 corresponding to the winding modes of closed strings. From a target space viewpoint, this 
 has led   to the development of double field theory (DFT) \cite{Siegel:1993xq,Hull:2009mi,Hohm:2010jy,Hohm:2010pp},
 which is currently under close scrutiny (see Refs. \cite{Aldazabal:2013sca,Berman:2013eva,Hohm:2013bwa}
  for reviews and a complete list of references).  On the other hand, our approach has a 3D/2D perspective 
  and does not use any results from DFT{\footnote{A new CFT approach of DFT was considered 
  recently in Refs. \cite{dftcft,dftcft2}. Previous work along this line includes Refs. \cite{Hull:2004in,Berman:2007xn,Dall'Agata:2008qz,Avramis:2009xi}.}}. Once more it is assumed that the 3D manifold has a 2D boundary, 
  and the consistency between the equations of motion and the boundary conditions uncovers an 
  extended set of relations that have to be satisfied. We comment on their relation to the flux formulation of DFT
  \cite{dftflux1,dftflux2,dftflux3,dftflux4} (see also \cite{fluxa,fluxb,fluxd,fluxc}). Finally we write down in very explicit terms a 3D sigma model 
  with doubled target where (i) all types of fluxes appear, (ii) 2-form and 2-vector deformations 
  are neither vanishing nor inverse of one another and (iii) the $R$-flux is \emph{not} of the type 
  that can be reduced to a \emph{standard} 2D sigma model. These properties suggest that this example is 
  a nontrivial toy model of a genuinely non-geometric background.

\section{Courant algebroids as protobialgebroids}\label{CA}

In this section we define and construct Courant algebroids 
that accommodate 3-index twists of any type, which will be 
identified with geometric and non-geometric fluxes that appear in string theory. We call such structures ``Courant-Roytenberg''
algebroids, since they were introduced in Ref. \cite{Roy}. Our approach in the presentation is to provide some basic definitions first, 
then apply them to construct a class of cases interesting for string theory, and finally to present in detail an explicit 
example. This approach will be followed in the following sections too.

\subsection{Definitions of bialgebroids}

Let us first define the notion of a protobialgebroid (PBA) and then discuss some particular limits. Note that PBAs are usually
defined using supermanifolds \cite{Roy}, but here we will use a more conservative, ``bosonic'' definition, which is more handy
 for applications in string theory (see however Ref. \cite{Deser:2014mxa}).

\begin{defn}\label{pba}
  Consider two dual vector bundles ($L,L^{\star}$) over a manifold M, equipped with the 
 following data:
\begin{itemize}
\item Skew-symmetric brackets $[\cdot,\cdot]_{L}$ on 
 $L$ and $[\cdot,\cdot]_{L^{\star}}$ on $L^{\star}$.
\item Bundle morphisms (anchors) $\rho:L\to\text{TM}$ and $\rho_{\star}:L^{\star}\to\text{TM}$.
\item Generalized 3-forms $\phi\in\G(\w^3L^{\star})$ and $\psi\in\G(\w^3L)$.
\end{itemize}
This structure is a \textbf{protobialgebroid} provided that for $X,Y,Z\in\G(L)$ and $\eta,\xi,\o\in\G(L^{\star})$ the following 
properties hold:
\begin{enumerate}
\item $[X,fY]_L=f[X,Y]_L+(\rho(X)f)Y$ and $[\eta,f\xi]_{L^{\star}}=f[\eta,\xi]_{L^{\star}}+(\rho_{\star}(\eta)f)\xi$~, 
      $f\in C^{\infty}(M)$~,
\item $\rho([X,Y]_L)=[\rho(X),\rho(Y)]_{\text{Lie}}+\rho_{\star}\phi(X,Y,\cdot)$ 
      and $\rho_{\star}([\eta,\xi]_{L^{\star}})=[\rho_{\star}(\eta),\rho_{\star}(\xi)]_{\text{Lie}}+\rho\psi(\eta,\xi,\cdot)$~,
\item $[[X,Y]_L,Z]_L+\text{c.p.}=\dd_{L^{\star}}\phi(X,Y,Z)+
      \phi(\dd_{L^{\star}}X,Y,Z)+\phi(X,\dd_{L^{\star}}Y,Z)+\phi(X,Y,\dd_{L^{\star}}Z)$~,
      $[[\eta,\xi]_{L^{\star}},\o]_{L^{\star}}+\text{c.p.}=\dd_{L}\psi(\eta,\xi,\o)+
      \psi(\dd_{L}\eta,\xi,\o)+\psi(\eta,\dd_{L}\xi,\o)+\psi(\eta,\xi,\dd_{L}\o)$.
\item $\dd_{L}\phi=0$ and $\dd_{L^{\star}}\psi=0$.
\end{enumerate}
\end{defn}

Although this definition does not appear as such in the literature{\footnote{It is briefly mentioned in Ref. \cite{KS} though.}}, it is just 
the appropriate generalization of the definition 3.8.3 for a quasi-Lie bialgebroid in the first reference of \cite{Roy}. 
The four enumerated properties in definition \ref{pba} are generalizations of the familiar 
properties of the tangent bundle. They are lifted to the 
general vector bundles $L$ and $L^{\star}$ with the aid of the maps $\rho$ and $\rho_{\star}$, called anchors. 
More precisely, the first property is just the Leibniz rule for each bundle. Recall that the tangent bundle,
whose sections are ordinary vector fields, is equipped with the standard Lie bracket of vector fields that satisfies the Leibniz 
rule
\be
[X,fY]_{\text{Lie}}=f[X,Y]_{\text{Lie}}+(Xf)Y~,
\ee
when $X,Y\in\G(\tb)$. It is then evident that property 1 
for each bundle is the direct generalization of this rule. 
The second property is a twisted version of $\rho$ and $\rho_{\star}$ being homomorphisms; 
$\rho$ is a $\phi$-homomorphism and $\rho_{\star}$ a $\psi$-homomorphism. The third property is a twisted version of the 
Jacobi identity. Finally, the fourth property states that 
the 3-objects $\phi$ and $\psi$ are closed with respect to 
the corresponding derivations on each vector bundle. These derivations are in turn 
the direct generalizations of the standard exterior derivative on the tangent bundle, which acts on $p$-forms  
raising their degree by one. In particular, they are 
simply defined as maps $\dd_L:\G(\w^p L^{\star})\to
\G(\w^{p+1}L^{\star})$ and $\dd_{L^{\star}}:\G(\w^p L)\to
\G(\w^{p+1}L)$, acting as follows \cite{MckenzieXu}:  
\bea\label{derivation}
\dd_{L} \omega(X_1,\dots ,X_{p+1})&=&\sum_{i=1}^{p+1}(-1)^{i+1}
\rho(X_i)\omega(X_1,\dots,\hat X_i,\dots ,X_{p+1})+
\nn\\ 
&&+\sum_{i<j}(-1)^{i+j}\omega([X_i,X_j]_{L},X_1,\dots ,\hat X_i,\dots ,\hat X_j, \dots ,X_{p+1})~,\\
\dd_{L^{\star}} \Omega(\eta_1,\dots ,\eta_{p+1})&=&\sum_{i=1}^{p+1}(-1)^{i+1}
\rho_{\star}(\eta_i)\Omega(\eta_1,\dots,\hat \eta_i,\dots ,\eta_{p+1})+\nn\\ 
\label{derivation2} &&+\sum_{i<j}(-1)^{i+j}\Omega([\eta_i,\eta_j]_{L^{\star}},\eta_1,\dots ,\hat \eta_i,\dots ,\hat \eta_j, \dots ,\eta_{p+1})~,
\eea
for arbitrary generalized $p$-forms $\omega\in\G(\w^p L^{\star})$ and $\Omega\in \G(\w^pL)$. The property 4 is essentially the 
set of Bianchi identities for the structure.

Given the general structure of a PBA, there are few 
special cases, depending on the presence or absence of the generalized 3-forms $\phi$ and $\psi$.
They are collected in the following table:
\begin{center}\boxed{
\begin{tabular}{ccc}
 \underline{$\phi$} & \underline{$\psi$} & \underline{Structure} 
 \\[4pt]
 $\ne 0$ & $\ne 0$ & Protobialgebroid
 \\[4pt]
 $\ne 0$ & $=0$ & Quasi-Lie bialgebroid
 \\[4pt]
 $=0$ & $\ne 0$ & Lie quasibialgebroid
 \\[4pt]
 $=0$ & $=0$ &  Lie bialgebroid
\end{tabular}}
\end{center} 

It is known from Ref. \cite{wein} that a Lie bialgebroid (LBA) gives rise to a Courant algebroid (CA) with vector bundle $E=L\oplus L^{\star}$, which will be defined below. 
More generally any PBA gives rise to a CA, as shown in Ref. \cite{Roy}.

\subsection{Protobialgebroids made explicit}

Let us construct a class of PBAs, based on nilmanifolds of step 2. Such manifolds are called ``twisted tori'' in the 
physics literature and they can be described as fibrations of toroidal fibers over toroidal bases, 
 whose tangent bundle can be obtained from the tangent bundle of standard tori by an appropriate deformation of 
degree $(1,1)$. Additionally we consider deformations by elements of degree $(0,2)$ and $(2,0)$, as explained below.  
This structure was already partially constructed in Ref.
\cite{Chatzistavrakidis:2013wra},
where it was called ``a QLBA with anchor on $L$ instead of TM''. The issue of anchors that did not project the 
vector bundles on $\tb$ 
will here be corrected using the more general protobialgebroid structure. 
Another difference is that in Ref. \cite{Chatzistavrakidis:2013wra} one had to invoke the 
Courant bracket as the bracket on the CA to do the computations, while now we are going to extract the 
consistent form 
of the bracket just from the deformation data, without a priori reference to the resulting CA. In particular, since the 
CA is not the standard one, the bracket on it is not simply the Courant bracket, but a more involved one, as explained in 
Ref. \cite{wein}. Finally, the associated CA was not constructed explicitly 
in Ref. \cite{Chatzistavrakidis:2013wra}, and we will 
complete this construction later in this section.

\subsubsection{Protobialgebroids over twisted tori}
\label{pbanil}

Consider M to be any $d$-dimensional step 2 nilmanifold with tangent and cotangent bundles spanned by
\bea
\theta_i&=&e_i^{a}(x)\partial_a~,
\\
e^i&=&e^i_{a}(x)\dd x^a,
~,
\eea
where early indices $a,b,\dots$ are curved (world indices) and late indices 
$i,j,\dots$ are flat (freely falling frame indices), and the frame components $e^i_a(x)$ are such that the Maurer-Cartan equations hold. 
We choose to work on a (generally curved) nilmanifold 
because it is a direct but nontrivial generalization of a flat torus. In particular, the tangent and cotangent bundles of 
any nilmanifold can be obtained from the toroidal ones by means of a $(1,1)$ deformation $h(x)=\sfrac 12h^b_a(x)\dd x^a\w \partial_b$:  
\be
\tb=e^{h(x)}\text{T}T^d~.
\ee
In the case of step 2 nilmanifolds the deformation can always be written as 
\be
h(x)=f^c_{ab}x^a\dd x^b\w\partial_c~,
\ee
for appropriate parameters $f^c_{ab}$ corresponding to the structure constants of the associated nilpotent Lie algebra. 
One can simply jump to the toroidal case by 
setting $f$'s to zero. Moreover, the choice of step 2, which amounts to the structure constant relation
\be 
f^c_{ab}f^b_{de}=0~, \quad \text{no summation in $b$}~,
\ee
is made for simplicity, but the general step case can also 
be addressed with the same methods. In the general case the deformation $h(x)$ is a more involved polynomial expression, and higher 
order relations among the structure constants hold (see Ref. \cite{Chatzistavrakidis:2014tda} for details).
The basis 1-vectors and 1-forms are dual,  
\be\label{pairing1}
\theta_i(e^j)=\d_i^j~,
\ee
reflecting the duality of $\tb$ and $\ctb$ as vector bundles.
Additionally, the 1-forms satisfy the usual Maurer-Cartan equations.

Let us endow the twisted torus with a (not necessarily closed) 2-form and a (not necessarily 
Poisson) 2-vector:
\bea 
B&=&\sfrac 12 B_{ij}e^i\w e^j~,\\
\beta&=&\sfrac 12\beta^{ij}\theta_i\w\theta_j~.
\eea 
Using $B$ and $\beta$ as deformations the tangent and cotangent bundles can be twisted accordingly{\footnote{These twisted bundles can 
equivalently be understood as graphs of $B$ and $\beta$ respectively, namely $L_B=\{X+B(X,\cdot);X\in \text{TM}\}$ and $L^{\star}_{\beta}=\{\eta+\beta(\eta,\cdot);\eta\in \ctb\}$}}:
\bea 
L_B&:=&e^{B}\text{TM}=\text{span}(\{\theta_i+B_{ij}e^j\})~,\\
L_{\beta}^{\star}&:=&e^{\beta}\text{T}^{\star}\text{M}=\text{span}(\{e^i+\beta^{ij}\theta_j\})~.
\eea 

Note that unlike $\tb$ and $\ctb$, which are dual due to the pairing \eqref{pairing1}, $L_B$ and $L_{\beta}^{\star}$ are not 
mutually dual{\footnote{Because the twist is not an element of O($d,d$).}}.
In Ref. \cite{Chatzistavrakidis:2013wra} we performed an independent 
change of basis on each bundle, such that duality is achieved. This is actually equivalent to performing an overall 
$e^Be^{\beta}$ twist{\footnote{The order of the twists, first with $\beta$ and then with $B$, counts. A twist 
of the form $e^{\beta}e^B$ would lead to another path, which is however equivalent for our purposes.}} on $\tb\oplus\ctb$. We simply get
\bea \label{Theta}
L_{B\beta}&=&e^Be^{\beta}\text{TM}=\text{span}(\{\Theta_i=\theta_i+B_{ij}e^j\})~,\\
L_{B\beta}^{\star}&=&e^Be^{\beta}\text{T}^{\star}\text{M}=\text{span}(\{E^i=e^i+\beta^{ik}B_{kj}e^j+\beta^{ij}\theta_j\})~.\label{E}
\eea 
We will never assume any constraining relation for $B\beta$; unlike what is usually assumed in the literature, here 
this combination is in general neither vanishing nor unity.
The pair that appears in the definition of the protobialgebroids that we consider is then $(L,L^{\star})=(L_{B\beta},L^{\star}_{B\beta})$.

According to the definition, we should specify elements $\phi\in\G(\w^{3}L^{\star}_{B\beta})$ and $\psi\in\G(\w^3L_{B\beta})$. 
In the spirit of 
twisting the tangent and cotangent bundle data, we consider arbitrary elements $H\in \G(\w^3\ctb)$ and
 $R\in \G(\w^3\tb)$ 
and twist them to give
\bea\label{phiexp}
\phi &=&\sfrac 16 \phi_{ijk}E^{i}\w E^j\w E^k\nn\\
&=&\sfrac 16\big( (1+\beta B)_{\rho}^i(1+\beta B)_{\sigma}^j(1+\beta B)_{\tau}^k \phi_{ijk}e^{\r}\w e^{\s}\w e^{\t}\nn\\
&&\quad+3(1+\beta B)_{\rho}^i
(1+\beta B)_{\sigma}^j\beta^{kl}\phi_{ijk}e^{\r}\w e^{\s}\w \theta_l\nn\\
&&\quad+3(1
+\beta B)_{\rho}^i\beta^{jl}\beta^{km}\phi_{ijk}e^{\r}\w \theta_{l}\w \theta_m
\nn\\
&&\quad +\beta^{il}\beta^{jm}\beta^{kn}\phi_{ijk}\theta_{l}\w \theta_{m}\w\theta_{n}\big)~,\\
\label{psiexp}
\psi &=& \sfrac 16 \psi^{ijk}\Theta_i\w \Theta_j\w\Theta_k\nn\\
&=&\sfrac 16\big(\psi^{ijk}\theta_{i}\w\theta_{j}\w\theta_{k}\nn\\
&&\quad+
3B_{kn}\psi^{ijk}\theta_{i}\w\theta_{j}\w e^n\nn\\
&&\quad +3B_{jm}B_{kn}\psi^{ijk}\theta_{i}\w e^{m}\w e^{n}\nn\\
&&\quad +B_{il}B_{jm}B_{kn}\psi^{ijk}e^{l}\w e^{m}\w e^{n}\big)~.
\eea
Eqs. \eqref{phiexp} and \eqref{psiexp} exhibit that 
in the presence of the twists ($\phi$ and $\psi$) and the deformations ($B$ and $\beta$)
there are all types of fluxes turned on, as they were identified e.g. in Ref.
\cite{Chatzistavrakidis:2013wra} in a less systematic way. 
Note that in some particular cases $H$ and $R$ can be identified with the derivations of $B$ and $\beta$ respectively.
 However, it will not always be the case in this paper that these identifications are made. This will be explicitly stated 
 when assumed.

Next we consider the bundle morphisms
\bea 
&& \rho:L_{B\beta}\to\tb~,\quad \rho(X)=e^{-\beta}e^{-B}X~,\label{anchor1}\\
&& \rho_{\star}:L_{B\beta}^{\star}\to \tb~, \quad \rho_{\star}(\eta)=\beta(e^{-\beta}e^{-B}\eta,\cdot)~.\label{anchor2}
\eea
Here and in the following we use the symbol $\beta$ also for the map $\beta:\ctb\to\tb$ (often denoted as $\beta^{\sharp}$ 
in the literature). These are the candidates for anchors, being the twisted versions of the corresponding anchors on $\tb$ (unit map) and $\ctb$ 
($\beta$-morphism).

Now we have to define skew-symmetric closed brackets on each of the two vector bundles.
Our strategy is once more to consider the corresponding
brackets on $\tb$ and $\ctb$ and twist them appropriately.
Let us use the notation $X,Y\in \G(L_{B\beta})$ and $\eta,\xi\in \G(L^{\star}_{B\beta})$.
Elements of $\tb$ are written as $\tilde X:=e^{-\beta}e^{-B}X$, and elements of $\ctb$ as $\tilde \eta:=e^{-\beta}e^{-B}\eta$. 
The bracket on $\tb$ is the standard Lie bracket of vector fields, while the bracket on $\ctb$ is 
\be 
[\tilde\eta,\tilde\xi]_{\text{K}}={\cal L}_{\beta(\tilde\eta,\cdot)}\tilde\xi-{\cal L}_{\beta(\tilde\xi,\cdot)}\tilde\eta-\dd\big(\beta(\tilde\eta,\tilde\xi)\big)~, \quad 
\tilde\eta,\tilde\xi \in \ctb~,
\ee
 $\dd$ being the standard de Rham differential. In the Poisson case this is the standard Koszul bracket
  of 1-forms.
 Then we consider the $e^{B}e^{\beta}$ twist of those brackets and write the Ans\"atze:
\bea\label{ansatz1}
[X,Y]_{L_{B\beta}}&=&e^Be^{\beta}[e^{-\beta}e^{-B}X,e^{-\beta}e^{-B}Y]_{\text{Lie}}
+V~,\\
{[}\eta,\xi]_{L^{\star}_{B\beta}}&=&e^{B}e^{\beta}[e^{-\beta}e^{-B}\eta,e^{-\beta}e^{-B}\xi]_{K}
+W~,\label{ansatz2}
\eea
where $V\in L_{B\beta}$ and $W\in L^{\star}_{B\beta}$ are associated to the twists and they should be determined by consistency with the 
definition \ref{pba}. 
In particular, for the bracket \eqref{ansatz1}, the second requirement of the definition, combined with the anchors 
defined above, gives
\bea
&&\rho([X,Y]_{L_{B\beta}})=[\rho(X),\rho(Y)]_{\text{Lie}}+\rho_{\star}\phi(X,Y,\cdot)\nn\\
\Leftrightarrow &&\rho(e^Be^{\beta}[e^{-\beta}e^{-B}X,e^{-\beta}e^{-B}Y]_{\text{Lie}})+
\rho(V)=[e^{-\beta}e^{-B}X,e^{-\beta}e^{-B}Y]_{\text{Lie}}
+\rho_{\star}\phi(X,Y,\cdot)\nn\\
\Leftrightarrow&&[e^{-\beta}e^{-B}X,e^{-\beta}e^{-B}Y]_{\text{Lie}}+\rho(V)=[e^{-\beta}e^{-B}X,e^{-\beta}e^{-B}Y]_{\text{Lie}}+
\rho_{\star}\phi(X,Y,\cdot)
\nn\\
\Leftrightarrow&&\rho(V)=\rho_{\star}\phi(X,Y,\cdot)\nn\\
\Leftrightarrow&&V=e^Be^{\beta}\beta(e^{-\beta}e^{-B}(\phi(X,Y,\cdot)),\cdot)~.
\eea
For the bracket \eqref{ansatz2} the analogous requirement is 
\bea \label{psihomo}
\rho_{\star}([\eta,\xi]_{L^{\star}_{B\beta}})=[\rho_{\star}(\eta),\rho_{\star}(\xi)]_{\text{Lie}}+\rho\psi(\eta,\xi,\cdot)~.
\eea
A similar computation leads to the result 
\be \label{lambda}
\rho_{\star}(W)=\beta(e^{-\beta}e^{-B}W,\cdot)=e^{-\beta}e^{-B}\psi(\eta,\xi,\cdot)-\sfrac 12 [\beta,\beta]_{\text{S}}(e^{-\beta}e^{-B}\eta,e^{-\beta}e^{-B}\xi,\cdot)~,
\ee
where $[\cdot,\cdot]_{\text{S}}$ is the Schouten bracket.
This equation should be solved for $W$ in order to fully determine the bracket on $L_{B\beta}^{\star}$. 
Unlike the previous case this is not straightforward, since it depends on the invertibility of $\rho_{\star}$ (while $\rho$ is 
always invertible in our approach). For invertible $\beta$, it is easy to solve for $W$ and plug it in the Ansatz \eqref{ansatz2}. 
However, the case of non-invertible $\beta$ is more interesting for our purposes. 
A way to solve \eqref{lambda} is to assume that the right hand side is zero, namely 
\be \label{psicondition}
\psi(\eta,\xi,\cdot)= \sfrac 12 e^Be^{\beta}[\beta,\beta]_{\text{S}}(e^{-\beta}e^{-B}\eta,e^{-\beta}e^{-B}\xi,\cdot)~,
\ee 
 and that 
$
\beta^3\phi=0.
$
Then we set
\be \label{lambda2}
W=\phi\big(e^Be^{\beta}\beta(e^{-\beta}e^{-B}\eta,\cdot),e^{B}e^{\beta}\beta(e^{-\beta}e^{-B}\xi,\cdot),\cdot\big)~.
\ee
We will see later that these conditions are mild enough to assure that nontrivial cases indeed exist.
According to the above, the brackets on the two vector bundles are determined to be{\footnote{
Note that these brackets seem to contain only $\phi$ and not $\psi$ explicitly. 
However, as it is clear from Eq. \eqref{psicondition}, $\psi$ is not zero and this is essential for 
the $\psi$-homomorphism equation \eqref{psihomo} to hold, as it should for a protobialgebroid. 
Moreover, $\psi$ will appear explicitly when we construct the bracket of the corresponding Courant algebroid, 
where the two twisted homomorphism conditions are replaced by a single homomorphism condition 
for the anchor of the Courant algebroid.}}
\bea\label{bra1}
&&\mkern-60mu[X,Y]_{L_{B\beta}}=e^Be^{\beta}\biggl([e^{-\beta}e^{-B}X,e^{-\beta}e^{-B}Y]_{\text{Lie}}
+\beta( e^{-\beta}e^{-B}(\phi(X,Y,\cdot)),\cdot)\biggl)~,\\
&&\mkern-60mu{[}\eta,\xi]_{L^{\star}_{B\beta}}=e^{B}e^{\beta}[e^{-\beta}e^{-B}\eta,e^{-\beta}e^{-B}\xi]_{\text{K}}
+\phi\big(e^Be^{\beta}\beta(e^{-\beta}e^{-B}\eta,\cdot),e^{B}e^{\beta}\beta(e^{-\beta}e^{-B}\xi,\cdot),\cdot\big)
~.\label{bra2}
\eea
The skew-symmetry of the brackets \eqref{bra1} and \eqref{bra2} follows from the skew-symmetry of the 
Lie and Koszul brackets and the antisymmetry of $\phi$. Closedness is also rather obvious. 
The big brackets in Eq. \eqref{bra1} contain an element of $\tb$. Then this element is 
acted upon with $e^Be^{\beta}$, yielding elements of $L_{B\beta}$, as required. Similarly, both terms in 
Eq. \eqref{bra2} are elements of $L_{B\beta}^{\star}$.
The brackets can be computed explicitly for the basis elements $\Theta_i$ and $E^i$; they yield the results
\bea 
[\Theta_i,\Theta_j]_{L_{B\beta}}&=&(f_{ij}^k-\beta^{km}\phi_{mij})\Theta_k~,
\\
{[}E^i,E^j]_{L^{\star}_{B\beta}}&=&(\theta_k\beta^{ij}-2\beta^{il}f^j_{lk}+\beta^{il}\beta^{jm}\phi_{klm})E^k~.
\eea

With the above elements $\phi$ and $\psi$, the brackets and the anchors, we have now collected all the input 
ingredients of a protobialgebroid, as required from the definition \ref{pba}. In the appendix we collect 
the proofs of the properties 1-4 in this definition.

\subsubsection{Explicit example}
\label{example}

In order to exhibit that nontrivial cases with nonvanishing $B$ and $\beta$ and with $B\beta\ne 1$ indeed exist, 
let us consider as an example the 3D nilmanifold based on the Heisenberg algebra with single structure constant $f^{3}_{\ 12}=1$. 
The full basis is 
\bea 
&&\theta_1=\partial_1~,\quad \theta_2=\partial_2+x^1\partial_3~,\quad \theta_3=\partial_3~,\\
&&e^1=\dd x^1~,\quad e^2=\dd x^2~,\quad e^3=\dd x^3-x^1\dd x^2~.
\eea
It can be checked that the manifold has a Poisson structure \cite{Rieffel}, given by the 2-vector
\be 
\theta_P=\m \theta_{1}\w\theta_{3}+\n \theta_2\w\theta_3~.
\ee
Therefore, any non-Poisson 2-vector will necessarily include $\theta_1\w\theta_2$. Here we consider such a 2-vector, 
\be 
\beta=\sqrt{c}\theta_1\w\theta_2~,
\ee
where $c$ is a real constant.
Its Schouten bracket gives:
\be
[\beta,\beta]_S=2R=2c\theta_{1}\w\theta_{2}\w\theta_{3}~,
\ee
where for this example we identified the Schouten bracket with $R$ (and $\dd B$ with $H$ below), which is not always the case.
Notably, $\beta$ being constant in the basis $\theta_i$ is enough to produce a nonvanishing 3-vector. 
Additionally we consider a 2-form proportional to the symplectic leaves of the manifold, which are $e^1\w e^3$ and $e^2\w e^3$. 
To be precise, we restrict the 2-form only on one leaf and take
\be 
B=Nx^1e^2\w e^3~.
\ee
This 2-form is not closed, giving
\be 
\dd B=H=Ne^{1}\w e^2\w e^3~.
\ee
The twisted bases are given as 
\bea 
L_{B\beta}&=&\text{span}(\{\Theta_i\}=\{\theta_1,\theta_2+Nx^1e^3,\theta_3-Nx^1e^2\})~,\\
L_{B\beta}^{\star}&=&\text{span}(\{E^i\}=\{e^1+\sqrt{c}Nx^1e^3+\sqrt{c}\theta_2,e^2-\sqrt{c}\theta_1,e^3\})~.
\eea

The closed brackets among the basis elements $\{\Theta_i,E^i\}$ are found via Eqs. (\ref{bra1}) and (\ref{bra2}). 
They are 
\bea
&&[\Theta_1,\Theta_2]_{L_{B\beta}}=\Theta_3~,\quad [\Theta_1,\Theta_3]_{L_{B\beta}}=-\sqrt{c}N\Theta_1~,\quad [\Theta_2,\Theta_3]_{L_{B\beta}}=-\sqrt{c}N\Theta_2~,\\
&&{[}E^1,E^2]_{L^{\star}_{B\beta}}= cNE^3~,\quad [E^1,E^3]_{L^{\star}_{B\beta}}=\sqrt{c}E^1~,\quad [E^2,E^3]_{L_{B\beta}^{\star}}=\sqrt{c}E^2~.
\eea
Note that these are different from the ones in Ref. \cite{Chatzistavrakidis:2013wra}, because the brackets have changed. 

We specify the anchors from Eqs. (\ref{anchor1}) and (\ref{anchor2}):
\bea\label{anchor1ex}
&&\rho(\Theta_i)=\theta_i
~,\\
&&\rho_{\star}(E^i)=\beta^{ij}\theta_j~.\label{anchor2ex}
\eea
Note that unlike Ref. \cite{Chatzistavrakidis:2013wra}, the 
anchors are morphisms to the $\tb$, as required. 

Finally, the 3-elements are:
\bea
\phi&=&Ne^{1}\w e^{2}\w e^{3}+\sqrt{c}N(e^{2}\w e^{3}\w \theta_2+e^{1}\w e^{3}\w\theta_1)+cNe^3\w \theta_{1}\w\theta_{2}~,\label{phiexample}
\\
\psi&=&c\theta_{1}\w\theta_{2}\w\theta_{3}+cNx^1(\theta_{3}\w\theta_{1}\w e^3+\theta_{2}\w\theta_{1}\w e^2)+c(Nx^1)^2\theta_1\w e^{2}\w e^{3}~.
\label{psiexample}\eea
It is simple to check that they satisfy the Bianchi identities $\dd_{L_{B\beta}}\phi=0$ and $\dd_{L^{\star}_{B\beta}}\psi=0$ respectively 
(see appendix).

\subsection{The induced Courant algebroid}

We recall the definition of a Courant algebroid according to Ref. \cite{wein}. 

\begin{defn}\label{CAdef}
 A \textbf{Courant algebroid} is a quadruplet $(E,[\cdot,\cdot]_{E},\langle\cdot,\cdot\rangle_{E},a)$ of the following data:
 \begin{itemize}
  \item a vector bundle $E$ over M,
  \item a skew-symmetric bracket on $\G(E)$,
  \item a non-degenerate 
symmetric bilinear form on $E$,
\item and an anchor map $a:
E\to \text{TM}$,
 \end{itemize}
 such that for $\mx_i\in \G(E)$:
\begin{enumerate}
\item $[[\mathfrak{X}_1,\mathfrak{X}_2]_E,\mathfrak{X}_3]_E+\text{c.p.}={\cal D}{\cal N}(\mathfrak{X}_1,\mathfrak{X}_2,\mathfrak{X}_3)$~,
\quad $3{\cal N}=\langle[\mathfrak{X}_1,\mathfrak{X}_2]_E,\mathfrak{X}_3\rangle_E+\text{c.p.}$~,
\item $a([\mx_1,\mx_2]_E)=[a(\mx_1),a(\mx_2)]_{\text{Lie}}$~,
\item $[\mx_1,f\mx_2]_E=f[\mx_1,\mx_2]_E+(a(\mx_1)f)\mx_2-\langle\mx_1,\mx_2\rangle_E{\cal D}f$~,~$f\in C^{\infty}(M)$~,
\item $\langle{\cal D}f,{\cal D}g\rangle_E=0$~, ~$f,g \in C^{\infty}(M)$~,
\item $a(\mx)\langle\mx_1,\mx_2\rangle_E=\langle[\mx,\mx_1]_E+{\cal D}\langle\mx,\mx_1\rangle_E,\mx_2\rangle_E+
\langle\mx_1,[\mx,\mx_2]_E+{\cal D}\langle\mx,\mx_2\rangle_E\rangle_E$~,
\end{enumerate}
where ${\cal D}:C^{\infty}(M)\to \G(E)$ is a map such that $\langle {\cal D}f,\mf{X}\rangle_E=\sfrac 12 a(\mf{X})f$~.
\end{defn}

 According to Roytenberg 
 there is a CA associated to any PBA \cite{Roy}. Its construction 
 is rather simple. Recall that according to Liu-Weinstein-Xu 
 the general bracket of a CA is not just the Courant bracket, but a more general expression \cite{wein}. The Courant bracket 
 only arises in 
 the case where the CA is standard, i.e. $E=\text{TM}\oplus 
 \text{T}^{\star}\text{M}$ {\textbf{and}} the bracket on 
 the cotangent bundle is taken to be zero (providing a trivial extension of the tangent bundle).
 In the case at hand the cotangent bundle is equipped with a non-trivial bracket, and 
 the CA is non-standard.
 Therefore the correct bracket on the CA should be 
 the $L_{B\beta}$ bracket plus the $L_{B\beta}^{\star}$ bracket with appropriate additional terms and twists.
 
 According to these, the vector bundle we consider is $E=L_{B\beta}\oplus L_{B\beta}^{\star}$,
with the bracket:
\bea 
[X+\eta,Y+\xi]_{E}&=&[X,Y]_{L_{B\beta}}+{\cal L}_X\xi-{\cal L}_Y\eta-\sfrac 12 
\dd_{L_{B\beta}} (X(\xi)-Y(\eta))\nn\\
&&+[\eta,\xi]_{L_{B\beta}^{\star}}+{\cal L}_{\eta}Y-{\cal L}_{\xi}X+\sfrac 12\dd_{L_{B\beta}^{\star}}(X(\xi)-Y(\eta))\nn\\
&&-\phi(X,Y,\cdot)-\psi(\eta,\xi,\cdot)~,
\eea
where the Lie derivatives are defined as 
\be 
{\cal L}_{X}=\dd_{L_{B\beta}}\iota_X+\iota_X\dd_{L_{B\beta}} \quad \text{and} \quad 
{\cal L}_{\eta}=\dd_{L^{\star}_{B\beta}}\iota_{\eta}+\iota_{\eta}\dd_{L^{\star}_{B\beta}}~.
\ee 
The anchor is just the sum of the two anchors,
\be 
a(X+\eta)=\rho(X)+\rho_{\star}(\eta)=e^{-\beta}e^{-B}X+\beta(e^{-\beta}e^{-B}\eta)~.
\ee
The symmetric bilinear is the standard one,
\be 
\langle X+\eta,Y+\xi\rangle_E=\sfrac 12(X(\xi)+Y(\eta))~.
\ee
These are the data of the CA that corresponds to the PBA structure of the previous sections. Note also that 
\be 
{\cal D}=\dd_{L_{B\beta}}+\dd_{L_{B\beta}^{\star}}~.
\ee 
It can be directly checked that the requirements 1-5 are satisfied. 
Notably, the anchor $a$ of the CA is a homomorphism due to property 2 in the definition \ref{CAdef}. Recall also that the 
maps $\rho$ and $\rho_{\star}$ are not exact homomorphisms, as dictated by property 2 in the definition \ref{pba}. 
This works as follows. Consider elements of $E$ which lie entirely in $L_{B\beta}$, i.e. $\mf{X}=X$ and $\eta=0$. 
The bracket of $E$ between such elements is
\be
[X,Y]_E=[X,Y]_{L_{B\beta}}-\phi(X,Y,\cdot)~.
\ee
Then we compute
\bea
a([X,Y]_E)&=&\rho([X,Y]_{L_{B\beta}})-\rho_{\star}\phi(X,Y,\cdot)\nn\\
&=&\big([\rho(X),\rho(Y)]_{\text{Lie}}+\rho_{\star}\phi(X,Y,\cdot)\big)-\rho_{\star}\phi(X,Y,\cdot)\nn\\
&=&[a(X),a(Y)]_{\text{Lie}}~,
\eea
as required. A similar computation holds for the dual case.

Although in this section we used only index-free notation, it is useful to introduce CA indices $I,J,\dots,$ ranging from 1 to $2d$. 
An arbitrary generalized vector is written as 
\be
\mathfrak{X}=(\mathfrak{X}_I)=(\mathfrak{X}^i,\mathfrak{X}_i)
\in\Gamma(E)~,
\ee
namely the index $I$ splits into upper and lower indices according to $\mathfrak{X}=\mathfrak{X}^i\Theta_i+\mathfrak{X}_iE^i$.

\section{The associated AKSZ sigma model}

\subsection{Topological sigma model, boundary terms and dynamics}

Every Courant algebroid has an associated (topological) sigma model of the type described by
Alexandrov, Kontsevich, Schwarz and Zaboronsky (AKSZ) in Ref. \cite{Alexandrov:1995kv}. This can be inferred e.g. by the discussion 
of Roytenberg in the paper \cite{Roytenberg:2006qz}. A physicists-friendly review is \cite{ikeda} (see also the paper \cite{Cattaneo:2009zx}).
The master action contains fields with ghost number 0, 1, 2 and 3. Let us focus 
on the 0-ghost sector of the action:
\be \label{aksz}
S_{\S_3}[X,A,F]=\int_{\S_3}\biggl(F_{a}\w\dd X^{a}+\sfrac 12 \eta_{IJ}A^I\w\dd A^J-P_{I}^aA^I\w F_a+\sfrac 16T_{IJK}A^I\w A^J\w A^K\biggl)~.
\ee
The explanation for the ingredients of this action is the following. This is a membrane topological action in 3D. The indices 
$I,J$ are Courant algebroid indices, while the index $a$ is a curved index, as before. $X^a$ are the world volume scalars on the membrane, or in other words 
the components of the map $X:\Sigma_3\to \text{M}$, M being the target spacetime. $A^I$ is valued in 
$\Omega^1(\S_3,X^{\star}E)$, where $X^{\star}$ denotes the pull back with respect to the world volume scalar fields.
 Additionally, 
$F_a$ is a world volume 2-form in $\Omega^2(\S_3,X^{\star}\text{T}^{\star}\text{M})$. In the membrane model it plays the role
of an auxiliary field that will be integrated out in the reduced string model. Moreover, $\eta$ is the O($d,d$) invariant metric, namely 
\be 
\eta_{IJ}=\begin{pmatrix}
0 & \one_d \\ \one_d & 0 
\end{pmatrix}~,
\ee
 and 
$P^a_I$ is the anchor matrix defined through the relation 
\be
a(\mathfrak{X}_I)=P_I^a(X)\partial_a~,
\ee
where $a:E\to \tb$ is the anchor of the CA.
Finally, $T\in \Omega^3(\S_3,X^{\star}E)$ is a generalized 3-form. 

We assume that the manifold $\S_3$ has a boundary, say $\partial\S_3:=\S_2$, since this is the relevant case for physical applications. The above action can be decorated with a general 
topological boundary term as in Ref. \cite{Halmagyi:2009te} (see also \cite{ikeda,Mylonas:2012pg}):
\be\label{boundarytop} 
S_{\partial\S_3,\text{top}}=\int_{\S_2}\sfrac 12 {\cal B}_{IJ}(X)A^I\w A^J~.
\ee
More explicitly, with the splitting $A^I=(q^i,p_i)$,
\be \label{tbt}
\sfrac 12{\cal B}_{IJ}(X)A^I\w A^J=\sfrac 12{\cal  B}_{ij}(X)q^i\w q^j+\sfrac 12{\cal B}^{ij}(X)p_i\w p_j
+\sfrac 12{\cal B}^i_j(X)q^j\w p_i~.
\ee
In the class of CAs we examine, all terms will play a role.

Additionally, in order to make contact with physics, dynamics should be added to the topological theory (thus breaking its topological 
nature). 
In this section our approach will be to study the 3D topological theory, then reduce it to the 
corresponding 2D field theory on the boundary and add dynamics at the level of this 2D theory. 
This is either done by simply adding a standard kinetic term 
\be 
\int_{\S_2} \sfrac 12 g_{ij}e^i\w\star e^j~,
\ee
or in certain cases a kinetic term formed with the inverse metric 
\be 
\int_{\S_2}\sfrac 12 g^{ij}p_i\w\star p_j~,
\ee
as in Ref. \cite{Halmagyi:2009te}.
The corresponding 2D theories are related to the dynamical sigma models discussed in Refs. \cite{Kotov:2004wz,Salnikov:2013pwa}. 

A final comment has to do with the functional dependence of the quantities that appear in the above 
actions. In this section we assume that the various background field components ${\cal B}_{IJ}$,
the anchor matrix $P^a_I$ and the twist $T$ solely depend on the scalar fields $X^a$.
These assumptions will be lifted in Section \ref{DFT}, where in the spirit of the first 
order formalism we will allow everything to depend both on $X^a$ and the corresponding momenta. 

\subsection{The AKSZ model for the Courant algebroid $E=L_{B\beta}\oplus L_{B\beta}^{\star}$}
\label{BbetaAKSZ}

Let us now specialize to the class of Courant algebroids that we discuss in this paper. The  ingredients of the 
topological membrane action \eqref{aksz} can be further specified. 
We hereby use the splitting $A^I=(q^i,p_i)$ referring to the basis $(e^i,\theta_i)$.
According to Eqs. (\ref{anchor1}-\ref{anchor2}), or more particularly Eqs. (\ref{anchor1ex}-\ref{anchor2ex}), we 
immediately obtain the components $P^a_I$ of the anchor matrix:
\bea
P^a_i&=&\mu e^a_i(X)~,\\
P^{ai}&=&\nu\beta^{ij}(X)e_j^a(X)~.
\eea
Note that we used the freedom to introduce parameters $\mu,\nu\in\{0,1\}$, since the CA structure is rigid against 
trivialization of the anchors. 
These parameters are relevant in taking interesting limits, as will become clear later in this section.

Given the above ingredients the bulk action is
\be
S_{\S_3}^{(\phi,\psi)}=\int_{\S_3}\biggl(F_{a}\w\dd X^{a}+ kq^i\w\dd p_i+k'p_i\w\dd q^i-
( \mu e_{i}^aq^i+\nu\beta^{ij}e_j^ap_i)\w F_a+f-\phi-\psi\biggl)~,
\label{bulkphipsi}\ee
with $f$ being the geometric flux
\be \label{f}
f=\sfrac 12 f^k_{ij}q^i\w q^j\w p_k~,
\ee
while, recalling that we work in the $(e^i,\theta_i)$ basis, $\phi$ and $\psi$ are the twists given by the expansions
\bea 
\label{phiexp2}
\phi &=&\sfrac 16\big( (1+\beta B)_{\rho}^i(1+\beta B)_{\sigma}^j(1+\beta B)_{\tau}^k\phi_{ijk}q^{\r}\w q^{\s}\w q^{\t}\nn\\
&+&3(1+\beta B)_{\rho}^i
(1+\beta B)_{\sigma}^j\beta^{kl}\phi_{ijk}q^{\r}\w q^{\s}\w p_l\nn\\
&+&3(1
+\beta B)_{\rho}^i\beta^{jl}\beta^{km}\phi_{ijk}q^{\r}\w p_{l}\w p_{m}\nn\\
&+&\beta^{il}\beta^{jm}\beta^{kn}\phi_{ijk}p_{l}\w p_{m}\w p_{n}\big)~,
\eea
and 
\bea
\label{psiexp2}
\psi &=& \sfrac 16\big(\psi^{ijk}p_{i}\w p_{j}\w p_{k}\nn\\
&+&
3B_{kn}\psi^{ijk}p_{i}\w p_{j}\w q^n \nn\\
&+&3B_{jm}B_{kn}\psi^{ijk}p_{i}\w q^{m} \w q^{n}\nn\\
&+&B_{il}B_{jm}B_{kn}\psi^{ijk}q^{l}\w q^{m}\w q^{n}\big)~.
\eea
We used the fact that we are free to introduce two additional parameters $k$ and $k'$. According to the general action 
\eqref{aksz} they have to satisfy 
\be \label{kk'}
k+k'=1~. 
\ee
The most symmetric choice is $k=k'=\sfrac 12$, and in absence of boundary one can always perform an 
integration by parts to change it to an arbitrary choice satisfying the condition \eqref{kk'}. 
In the presence of a 2D boundary these parameters 
are used for interpolation between different limits.

Now let us specify the boundary action. This contains all possible terms incorporating $B,\beta$ and $h$ 
deformations{\footnote{Although when we discuss specific examples we never add excess geometric flux on the twisted torus, 
in the general discussion such a possibility is retained.}}.
Accordingly, the boundary action is given by Eqs. \eqref{boundarytop} and \eqref{tbt},
with each set of components given as 
\bea 
{\cal B}_{ij}=B_{ij}
~,\quad
{\cal B}^{ij}=\beta^{ij}
~,\quad 
{\cal B}_i^j=h_i^j
~.
\eea
The full action that we consider is then 
\be 
S=S_{\S_3}^{(\phi,\psi)}+S_{\partial\S_3,\text{top}}^{(B,\beta,h)}~.
\ee
This action comes with a set of consistency conditions. First, the boundary conditions should match with the 
equations of motion on the boundary. This implies that we have to vary the action with respect to $X^a,q^i$ and $p_i$, 
set the variations to zero and determine appropriate boundary conditions. 
Performing this task we obtain 
\bea 
\d_{X^a}S|_{\S_2}&=&F_a+\sfrac 12\partial_a{B}_{jk}q^j\w q^k+\sfrac 12 \partial_a{\beta}^{jk}p_j\w p_k+
\sfrac 12\partial_a{h}_j^kq^j\w p_k=0~,
\nn\\
\d_{q^i}S|_{\S_2}&=&-(k'p_i+{B}_{ij}q^j+\sfrac 12{h}_i^jp_j)=0~,\nn\\
\d_{p_i}S|_{\S_2}&=&-(kq^i+{\beta}^{ij}p_j-\sfrac 12{h}^i_jq^j)=0~.\label{eoms}
\eea
These conditions are generalizations of the ones that appear e.g. in \cite{ikeda} and \cite{Hofman:2002rv} and 
they can be solved in many ways, as we will explore below.
The second consistency condition that has to be satisfied reads as
\bea \label{bbcondition}
(\mu e_{i}^a(X)q^i+\nu\beta^{ij}e_j^ap_i)\w F_a=f-\phi-\psi \qquad \text{on $\S_2$}~.
\eea
Normally this condition
 follows from the classical master equation \cite{ikeda}. Alternatively it can be viewed as 
 vanishing of the sector of the bulk action that does not reduce to the boundary via the 
 field equations.

Let us now explore some boundary conditions. First, we consider
\bea 
F_a|_{\S_2}&=&-\sfrac 12\partial_a{B}_{jk}q^j\w q^k-\sfrac 12 \partial_a{\beta}^{jk}p_j\w p_k-\sfrac 12\partial_a{h}_j^kq^j\w p_k~,
\nn\\
\d q^i|_{\S_2}&=&0~,\nn\\
(kq^i+{\beta}^{ij}p_j-\sfrac 12{h}^i_jq^j)|_{\S_2}&=&0~.\label{bc1}
\eea
Notably, the mild condition ${h}^i_k{h}^k_j=0$ allows us (for $k\ne 0$) to write
\be \label{qp}
q^i=-\sfrac 1{k}\chi_k^i{\beta}^{kj}p_j~,
\ee
where we introduced shorthand notation $\chi=1+\sfrac 1{2k}{h}$.
A medium long calculation shows that \eqref{bbcondition} reduces to the bulk/boundary consistency condition
\begin{empheq}[box=\fbox]{align}
{\cal R}^{ijk}
-\sfrac 1k {\cal Q}_n^{[ij}\beta^{\underline{p}k]}\chi^{n}_p
+\sfrac 1{k^2}{\cal F}^{[i}_{mn}\beta^{\underline{p}j}\beta^{\underline{q}k]}
\chi^{m}_p\chi^{n}_q
-\sfrac 1{k^3}{\cal H}_{lmn} \beta^{pi}\beta^{qj}\beta^{rk}
\chi^{l}_p\chi^{m}_q\chi^{n}_r=0~,
\label{condition1}
\end{empheq}
where we defined
\bea
{\cal R}^{ijk}&=&\psi^{ijk}-3\nu\beta^{[i\underline{l}}\theta_l{\beta}^{jk]}+\beta^{li}\beta^{mj}\beta^{nk}\phi_{lmn}~,\nn\\
{\cal Q}^{ij}_k&=&-3\mu \theta_k{\beta}^{ij}+3\nu\beta^{[i\underline{l}}\theta_l{h}^{j]}_k+3B_{lk}\psi^{ijl}
+3(1+\beta B)^l_k\beta^{mi}\beta^{nj}\phi_{lmn}~,\nn\\
{\cal F}_{jk}^i&=&-3\mu\theta_{[j}{h}_{k]}^{i}-3f^{i}_{jk}-3\nu\beta^{il}\theta_l{B}_{jk}+3B_{lj}B_{mk}\psi^{lmi}
+3(1+\beta B)^l_j(1+\beta B)^m_k\beta^{ni}\phi_{lmn}~,\nn\\
{\cal H}_{ijk}&=&(1+\beta B)^l_i(1+\beta B)^m_j(1+\beta B)^n_k\phi_{lmn}-3\mu\theta_{[i}{ B}_{jk]}
+B_{li}B_{mj}B_{nk}\psi^{lmn}~.\label{hfqr}
\eea
This long expression reveals the rich structure of the type of models we consider. 
In certain limits the condition \eqref{condition1} simplifies drastically and reduces to known results, as we will 
discuss in the next section.

Second, consider the alternative boundary conditions
\bea 
F_a|_{\S_2}&=&-\sfrac 12\partial_a{ B}_{jk}q^j\w q^k-\sfrac 12 \partial_a{\beta}^{jk}p_j\w p_k-\sfrac 12\partial_a{h}_j^kq^j\w p_k~,
\nn\\
(k'p_i+{B}_{ij}q^j+\sfrac 12{h}_i^jp_j)|_{\S_2}&=&0~,\nn\\
\d p_i|_{\S_2}&=&0~.\label{bc2}
\eea
As before, for $k'\ne0$ and defining $\chi'=1-\sfrac 1{2k'}{h}$ we can write
\be \label{pq}
p_i=-\sfrac 1{k'}\chi'^k_i{B}_{kj}q^j~,
\ee
which will now yield a consistency condition different from the previous case. The new calculation 
leads to
\begin{empheq}[box=\fbox]{align}
 {\cal H}_{ijk}-\sfrac 1{k'}{\cal F}_{[ij}^n
{B}_{\underline{p}k]}\chi'^p_n
+\sfrac 1{k'^2}{\cal Q}_{[i}^{mn}{B}_{\underline{p}j}B_{\underline{q}k]}\chi'^p_m\chi'^q_n
-\sfrac 1{k'^3}{\cal R}^{lmn}
{B}_{pi}{B}_{qj}{B}_{rk}\chi'^p_l\chi'^q_m\chi'^r_n=0~,
\label{condition2}
\end{empheq}
with the same definitions (\ref{hfqr}).

Finally, let us comment on the possibility of using the boundary conditions
\bea 
F_a|_{\S_2}&=&-\sfrac 12\partial_a{B}_{jk}q^j\w q^k-\sfrac 12 \partial_a{\beta}^{jk}p_j\w p_k-\sfrac 12\partial_a{h}_j^kq^j\w p_k~,
\nn\\
(k' p_i+{ B}_{ij}q^j+\sfrac 12{h}_i^jp_j)|_{\S_2}&=&0~,\nn\\
(kq^i+{\beta}^{ij}p_j-\sfrac 12{ h}^i_jq^j)|_{\S_2}&=&0~.
\eea
Now we obtain that both \eqref{qp} and \eqref{pq} should hold. This leads to the equations
\bea
&& \big(1-\sfrac 1{kk'}\chi'{B} \chi{\beta}\big)_i^j p_j=0~,\nn\\
&& \big(1-\sfrac 1{kk'}\chi{\beta}\chi'{ B} \big)^i_jq^j=0~
\eea
which in general force $p_i=q^i=0$, which is way too restrictive.

As a final remark, it should be clear that the above sets of boundary conditions are just two illustrative cases and they do not exhaust 
the range of possibilities,
since one can impose mixed boundary conditions 
too. We will encounter interesting cases of mixed boundary conditions later.

\subsection{Bulk/boundary versus integrability conditions for Dirac structures}
\label{BCvsIC}

Let us explore some limits of the bulk/boundary consistency conditions \eqref{condition1} and \eqref{condition2} and show that they reduce to previously obtained results. In particular we show that they are 
equivalent to the integrability conditions for twisted almost Dirac structures.
Recall that a Dirac structure $L$ is a subbundle of a CA $E$ which satisfies the following two conditions:
\bea 
\langle L,L\rangle_E&=&0~,\\
{[}L,L]_E&\in& L~,
\eea
namely it is maximally isotropic and involutive with respect to the CA bracket \cite{dirac}. 
An almost Dirac structure is just a maximal isotropic subbundle, i.e. the bundle before the second condition is imposed. 
Imposing the closure condition yields an integrability condition for $L$. 

In Ref. \cite{Chatzistavrakidis:2013wra}, the study of twisted almost Dirac structures led us to the results summarized in 
Table \ref{table1}
for the vector bundles $L_B=e^{B}\tb$ and $L_{\beta}^{\star}=e^{\beta}\ctb$ with various choices of the CA 
bracket{\footnote{Slight differences to \cite{Chatzistavrakidis:2013wra} in factors and signs are due to change of conventions on one hand
and different way 
of presentation on the other hand.}}.
\begin{center}
\begin{table}[h]\boxed{
\begin{tabular}{ccc}
 \underline{Twisted Dirac structure} & \underline{Bracket $[\cdot,\cdot]_T$} & \underline{Condition} 
 \\[4pt]
 $L_B$ & $[\cdot,\cdot]_H$ & $\dd B=H$
 \\[4pt]
 $L^{\ast}_{\beta}$ & $[\cdot,\cdot]_R$ & $\sfrac 12 [\beta,\beta]=R$ 
 \\[4pt]
 $L_B$ & $[\cdot,\cdot]_R$ & $(\dd B)_{ijk}-\sfrac 13 B_{il} B_{jm} B_{kn}R^{lmn}=0$
 \\[4pt]
 $L^{\ast}_{\beta}$ & $[\cdot,\cdot]_H$ &  $([\beta,\beta])^{ijk}-\sfrac 23\beta^{il}\beta^{jm}\beta^{kn}H_{lmn}=0$
\\[4pt]
$L_B$ & $[\cdot,\cdot]_{HR}$ & $\dd B=H$ and  $B_{il}B_{jm}B_{kn}R^{lmn}=0$
\\[4pt]
$L^{\ast}_{\beta}$ & $[\cdot,\cdot]_{HR}$ & $\sfrac 12 [\beta,\beta]=R$ and $\beta^{il}\beta^{jm}\beta^{kn}H_{lmn}=0$
\end{tabular}
}\caption{Integrability conditions for the almost Dirac structures $L_{B}$ and $L_{\beta}^{\star}$ with $H$ or/and $R$ twists.}
\label{table1}
\end{table}
\end{center}
The brackets that appear on the table are the Courant bracket twisted by $H$, $R$ or both{\footnote{One should be cautious about the 
differences with the twists $\phi$ and $\psi$. In Ref. \cite{Chatzistavrakidis:2013wra} it was assumed that the bracket twists are 
exactly $H=\dd B$ and $R=\sfrac 12[\beta,\beta]_{\text{S}}$, while in the present setting the twists $\phi$ and $\psi$ are more general. We clarify 
this further below.}}.
With the choice of bracket $[\cdot,\cdot]_T$, the integrability condition of the third column must hold. 
The conditions in the first and second row are of course standard. Additionally, the condition in the fourth row is also standard 
and it  corresponds to the $H$-twisted Poisson sigma model
\cite{psm,sw,severa}.
This table can also be obtained in the context of the AKSZ sigma models and we now show how (see also \cite{Hofman:2002rv,ikeda} 
for related discussions). 

\paragraph{Case 1: Dirac structure $L_{B}$.}
In this case we set $\beta=0$ and $h=0$, and we keep only a nonvanishing $B$. Additionally, we make the following choice of parameters:
\be
k=0~,\quad k'=1~,\quad \mu=1~,\quad  \nu=\{0,1\}~.
\ee
The relation \eqref{pq} simply becomes
\be \label{pqlb}
p_i=-B_{ij}q^j~.
\ee
Then Eq. \eqref{hfqr} gives
\bea 
{\cal R}^{ijk}&=&\psi^{ijk}~,\nn\\
{\cal Q}^{ij}_k&=&3B_{lk}\psi^{ijl}~,\nn\\
{\cal F}^i_{jk}&=&-3f^i_{jk}+3B_{lj}B_{mk}\psi^{lmi}~,\nn\\
{\cal H}_{ijk}&=&\phi_{ijk}-3\theta_{[i}B_{jk]}+B_{li}B_{mj}B_{nk}\psi^{lmn}~.
\eea
The bulk/boundary consistency condition \eqref{condition2} reduces to the significantly simpler expression
\bea
\phi_{ijk}-3\theta_{[i}B_{jk]}-3f^l_{[ij}B_{k]l}=0~,
\eea
or equivalently 
\begin{empheq}[box=\fbox]{align}
\phi-\dd B=0
\label{iclb}
\end{empheq}
in index-free notation.
In order to  compare this condition with the ones given in Table \ref{table1}, we recall that these results  were obtained by twisting the bracket on $L_B$ with a 3-vector $R$ and/or a 3-form $H$.
Therefore it is useful to write the fluxes $\phi$ and $\psi$ reduced to the boundary as
\be
\phi+\psi=\sfrac 16 (\phi_{ijk}+B_{li}B_{mj}B_{nk}\psi^{lmn})q^i\w q^j\w q^k+\sfrac 16 \psi^{ijk}p_i\w p_j\w p_k~.
\ee
The first line in  Table \ref{table1} corresponds to the case of $\phi_{ijk}=H_{ijk}$ 
and $\psi^{ijk}=R^{ijk}=0$, whence the integrability condition (\ref{iclb}) gives $dB=H$. 
Similarly, the third line corresponds to $\phi_{ijk}=B_{il}B_{jm}B_{kn}\psi^{lmn}$ and $\psi^{ijk}=R^{ijk}$,  leading to the integrability condition 
$(dB)_{ijk}=\sfrac 13 B_{il}B_{jm}B_{kn}R^{lmn}$.
Finally, for the fifth line we have $\phi_{ijk}=H_{ijk}$ and $\psi^{ijk}=R^{ijk}$ thus giving $dB=H$ and
$B_{il}B_{jm}B_{kn}R^{lmn}=0$.

\paragraph{Case 2: Dirac structure $L_{\beta}^{\star}$.}
In this case we set $B=0$ and $h=0$, and we keep only a nonvanishing $\beta$. 
We choose the parameters 
\be 
k=1~,\quad k'=0~,\quad \mu=1~,\quad \nu=0~.
\ee 
The relation \eqref{qp} becomes
\be \label{qpbeta}
q^i=-\beta^{ij}p_j~,
\ee
and the definitions \eqref{hfqr}
\bea
{\cal R}^{ijk}&=&\psi^{ijk}+\beta^{li}\beta^{mj}\beta^{nk}\phi_{lmn}~,\nn\\
{\cal Q}^{ij}_k&=&-3\theta_k\beta^{ij}+3\beta^{li}\beta^{mj}\phi_{klm}~,\nn\\
{\cal F}^i_{jk}&=&-3f^i_{jk}+3\beta^{li}\phi_{jkl}~,\nn\\
{\cal H}_{ijk}&=&\phi_{ijk}~.
\eea
Then the bulk/boundary consistency condition \eqref{condition1} reduces to
\bea
\psi^{ijk}-3\beta^{[i\underline{l}}\theta_l\beta^{jk]}-3f^i_{mn}\beta^{mj}\beta^{nk}=0~,
\eea
or equivalently
\begin{empheq}[box=\fbox]{align} \label{bblbeta}
\psi-\sfrac 12[\beta,\beta]_{\text{S}}=0~.
\end{empheq}
As before, this expression directly yields the integrability conditions for the almost Dirac structure $L_{\beta}^{\star}$ 
appearing in the second, fourth and sixth rows of Table \ref{table1}. To make this explicit we write  the fluxes $\phi$ and $\psi$ reduced to the boundary as
\be
\phi+\psi=\sfrac 16 \phi_{ijk}q^i\w q^j\w q^k+\sfrac 16 (\psi^{ijk}+\beta^{li}\beta^{mj}\beta^{nk}\phi_{lmn})p_i\w p_j\w p_k~,
\ee
Then the second row in  Table \ref{table1} corresponds to $\phi_{ijk}=0$ and $\psi^{ijk}=R^{ijk}$, thus reducing (\ref{bblbeta}) to $R=\sfrac 12[\beta,\beta]_{\text{S}}$. 
Similarly, the fourth line in the table
is obtained when $\phi_{ijk}=H_{ijk}$ and $\psi^{ijk}=-\beta^{li}\beta^{mj}\beta^{nk}\phi_{lmn}$, resulting in the integrability condition $\beta^{il}\beta^{jm}\beta^{kn}H_{lmn}=\sfrac 32([\beta,\beta]_{\text{S}})^{ijk}$.
For the sixth line $\phi_{ijk}=H_{ijk}$ and $\psi^{ijk}=R^{ijk}$ and the integrability condition (\ref{bblbeta}) reduces to
$R=\sfrac 12[\beta,\beta]_{\text{S}}$ and $\beta^{li}\beta^{mj}\beta^{nk}H_{lmn}=0$.

The pattern is already obvious. 
The choice of bracket corresponds to the choice of twist in the membrane 
action. The choice of Dirac structure deformation corresponds to the choice of boundary condition on the boundary string. The integrability 
condition corresponds to consistency of the boundary conditions with the bulk action. This dictionary is summarized as:
\begin{center}\boxed{
\begin{tabular}{cc}
 \underline{Courant algebroid} & \underline{Sigma model}
 \\[4pt]
 Bracket twist $[\cdot,\cdot]_T$ & Bulk term $-\int_{\S_3}T $
 \\[4pt]
 Dirac structure deformation $L_{\cal B}$ & Boundary term $\int_{\partial\S_3}{\cal B}$
 \\[4pt]
 Integrability condition for Dirac structure & Bulk/boundary consistency condition
 \\[4pt]
\end{tabular}}
\end{center}

\subsection{2D sigma models with dynamics}

Up to now we discussed the 3D topological field theory. For physical applications, notably for string theory, it is 
necessary to look at the corresponding 2D theory on the boundary and add dynamics to it. Let us first revisit the two cases of 
Section \ref{BCvsIC} from this perspective. 

For the first case of $L_{B}$, the $F_a$ equation of motion yields
\be \label{qe}
q^i=e^i_a\dd X^a=e^i~.
\ee
Using this to integrate out the auxiliary field and adding dynamics in the standard way, we obtain the familiar 2D field theory
with Wess-Zumino term
\bea 
S=\int_{\S_2}\big(\sfrac 12 g_{ij}e^i\w\star e^j+\sfrac 12 B_{ij}e^i\w e^j\big)-
\int_{\S_3}\sfrac 16 \phi_{ijk}e^i\w e^j\w e^k~.
\label{gB}
\eea
In the second case of $L_{\beta}^{\star}$, the $F_a$ equation again gives (\ref{qe}), and integrating out the auxiliary 2-form 
produces the action
\be \label{wzpsm}
S=\int_{\S_2}\big(\sfrac 12 \tilde g^{ij} p_i\w\star p_j+ p_i\w e^i+\sfrac 12 \beta^{ij}p_i\w p_j\big)
-\int_{\S_3}\sfrac 16\psi^{ijk}p_{i}\w\ p_j\w p_k~,
\ee
where in the spirit of the first order formalism we added dynamics with the inverse 
metric{\footnote{More precisely this is \emph{not} an inverse metric but the standard metric on the dual algebroid structure.}}
$\tilde g^{ij}$, exactly as in Ref. \cite{Halmagyi:2009te}. 
Some remarks regarding the action (\ref{wzpsm}) are in order. First,
 when the bracket is twisted only with a 3-form $H$, one has $\psi^{ijk}=-\beta^{li}\beta^{mj}\beta^{nk}H_{lmn}$ 
and this is precisely the $H$-twisted Poisson sigma model \cite{psm}
on a nilmanifold. In that case one can write the standard kinetic term. 
Furthermore, if $\beta$ is invertible and its inverse is equal to $B$, then 
the action (\ref{wzpsm}) with $g=-B\tilde g^{-1}B$ is equivalent to (\ref{gB}). 

\subsection{Explicit sigma model with both $B$ and $\beta$, and $B\beta\notin \{0,1\}$}
\label{akszexplicit}

In Section \ref{BCvsIC} we showed that the general formulae of Section \ref{BbetaAKSZ} reproduce known results in the limits 
 $B=0$ and $\beta=0$ respectively. However, in general none of $B$ and $\beta$ is zero, and moreover they do not have to satisfy 
 any relation of the sort $B\beta=1$, as is sometimes assumed. The results of Section  \ref{BbetaAKSZ} reflect such 
 general cases. In the present section we want to show that these results are not empty, in the sense that 
 there indeed exist nontrivial cases where the consistency conditions of the AKSZ sigma model can be satisfied. 
 
 In order to be very explicit, let us consider the toy example of Section \ref{example}, where the nonvanishing components 
 of $B$ and $\beta$ are $B_{23}=NX^1$ and $\beta^{12}=\sqrt{c}$. Therefore
 \be
 B\beta=\begin{pmatrix}
         0 & 0 & 0  \\ 0 & 0 & 0 
          \\ \sqrt{c}NX^1 & 0 & 0
        \end{pmatrix}~,
 \ee
 which is neither vanishing nor unity. In very explicit terms, the sigma model is 
 \bea 
 S=\int_{\S_3}&&\biggl(F_a\w\dd X^a+\sfrac 12 q^i\w\dd p_i+\sfrac 12 p_i\w\dd q^i-(q^1-\sqrt{c}p_2)\w F_1-(q^2+\sqrt{c}p_1)\w F_2
 \nn\\ &&-
 (q^3+X^1q^2+\sqrt{c}X^1p_1)\w F_3 +q^1\w q^2\w p_3 -Nq^1\w q^2\w q^3\nn\\
 &&-\sqrt{c}N(q^2\w q^3\w p_2+q^1\w q^3\w p_1)-cNq^3\w p_1\w p_2-
 cp_1\w p_2\w p_3\nn\\
 &&-cNX^1(p_3\w p_1\w q^3+p_2\w p_1\w q^2)-c(NX^1)^2p_1\w q^2\w q^3\biggl) \nn\\
 +\int_{\S_2}&&\biggl(NX^1q^2\w q^3+\sqrt{c}p_1\w p_2\biggl)~,
 \eea
 where the indices $a$ and $i$ run from 1 to 3, and we have made the choices $k=k'=\sfrac 12$ and $\mu=\nu=1$. 
 Proceeding with the variations, the $\d_{X^a}$ ones directly lead to the boundary condition
 \be \label{Fexample}
 F_1=-Nq^2\w q^3~,\quad F_2=F_3=0~.
 \ee
 The variations $\d_{p_i}$ and $\d_{q^i}$ lead to the following set of relations:
 \bea
&& (\sfrac 12 q^1+\sqrt{c}p_2)\d p_1=0~,\quad  (\sfrac 12 q^2-\sqrt{c}p_1)\d p_2=0~,\quad  (\sfrac 12 q^3)\d p_3=0~,\nn\\ 
&& (\sfrac 12p_1)\d q^1=0~, \quad (\sfrac 12 p_2+NX^1q^3)\d q^2=0~, \quad  (\sfrac 12 p_3-NX^1q^2)\d q^3=0~.\label{varsexample}
 \eea
Additionally, taking into account (\ref{Fexample}),  the bulk/boundary consistency condition is 
\bea 
&& N(q^1-\sqrt{c}p_2)\w q^2\w q^3+q^1\w q^2\w p_3 -Nq^1\w q^2\w q^3\nn\\
 &&-\sqrt{c}N(q^2\w q^3\w p_2+q^1\w q^3\w p_1)-cNq^3\w p_1\w p_2-
 cp_1\w p_2\w p_3\nn\\
 &&-cNX^1(p_3\w p_1\w q^3+p_2\w p_1\w q^2)-c(NX^1)^2p_1\w q^2\w q^3=0~. \label{bbexample}
\eea
Now we have to choose appropriate boundary conditions, consistent with Eqs. (\ref{varsexample}) and (\ref{bbexample}). 
The choice corresponding to (\ref{bc2}) is
\be
\d p_i=0~,\quad p_1=0~,\quad p_2=-2NX^1q^3~,\quad p_3=2NX^1q^2~.
\ee
It is observed that Eq. (\ref{psiexp2}) gives $\psi=0$. This is a legitimate possibility but it is not so interesting because 
it makes one of the twists vanish. 
On the other hand, the choice $\d q^i=0$ of (\ref{bc1}) is \emph{not} consistent with (\ref{bbexample}) for 
$c\ne 0$. 
This indicates that mixed boundary conditions are appropriate  in order to keep both $\phi$ and $\psi$ nonvanishing. 
We can find such conditions by first noting that 
\begin{itemize}
 \item $\d q^1\ne 0 \quad \Rightarrow\quad p_1=0 \quad \Rightarrow \quad \psi=0$~,
 \item $\d p_3\ne 0\quad\Rightarrow\quad q^3=0\quad\Rightarrow\quad \phi=0$~,
 \item $\big(\d p_1\ne 0 \quad \text{and}\quad  \d q^2\ne 0\big) \quad \Rightarrow \quad \big(p_2\propto q^3\quad \text{and} 
 \quad q^1\propto q^3\big)\quad \Rightarrow \quad \phi=0$~,
 \item $ \big(\d p_2\ne 0 \quad \text{and}\quad  \d q^3\ne 0\big) \quad \Rightarrow \quad \big(p_3\propto p_1\quad \text{and} 
 \quad q^2\propto p_1\big)\quad \Rightarrow \quad \psi=0$~.
\end{itemize}
This leads to the necessary requirements
\be
\d q^1=0~,\quad \d p_3=0~,\quad \big(\d p_1=0 \quad \text{or}\quad \d q^2=0\big)~, 
\quad \big(\d p_2=0 \quad \text{or}\quad \d q^3=0\big)~. 
\ee
Let us choose $\d q^2=\d p_2=0$ for the last two requirements. 
The remaining boundary conditions  from (\ref{varsexample}) are
\be 
q^1=-2\sqrt{c}p_2~, \quad p_3=2NX^1q^2~,\quad \text{on $\S_2$}~.
\ee
In order to be able to solve the bulk/boundary consistency condition we choose additionally 
\be
q^3+ \sfrac 12 \sqrt{c}X^1p_1=0 \quad \text{on $\S_2$}~.
\ee
Then we find that on the boundary
\bea 
\phi&=&\sfrac 12 cN X^1q^2\w p_1\w p_2~,\\
\psi&=&2\sqrt{c}Nq^2\w q^3\w p_2~,
\eea 
and it is checked that the condition (\ref{bbexample}) is satisfied. This shows that the boundary conditions that were chosen 
are consistent with the AKSZ action, while both twists $\phi$ and $\psi$ and both deformations $B$ and $\beta$ are nonvanishing. 
Focusing on 2D, the corresponding action can be brought to the form
\bea 
\int_{\S_2}\big(\sfrac 12 g_{ij}e^i\w\star e^j+\sfrac 12 p_i\w e^i+NX^1e^2\w e^3+\sqrt{c}p_1\w p_2-\sqrt{c}NX^1p_1\w e^3\big)~.
\eea
This is a nontrivial case from the general class of 2D field theories called Dirac sigma models, 
introduced and studied in Refs. \cite{Kotov:2004wz,Salnikov:2013pwa}.

\section{Toward a sigma model description  of  double field theory}
\label{DFT}

We would like to examine to what extend the approach we adopted up to now can be carried on to 
account for genuinely non-geometric cases.  As mentioned in the introduction,
non-geometric situations are better understood in the 
doubled formalism, where non-geometry is triggered by the presence of dual coordinates. In the doubled field theory these where implemented in an 
effective field theory on some doubled spacetime.
Here we do not work in a target space field theory framework, but instead we formulate the appropriate sigma model. 
This is close in spirit to the inspiring attempt of Ref. \cite{Mylonas:2012pg} to describe non-geometric backgrounds in the 
context of AKSZ sigma models. The authors used this approach to discuss quantization of non-geometric backgrounds and  limited their description to the single presence of $R$ flux.
This case is however known to
be T-dual to standard $H$ flux backgrounds and as such it is \emph{not} a genuinely non-geometric background. In the following 
we will extend and generalize the scope of
  AKSZ  inspired sigma models to account for more general cases.

\subsection{Sigma models with doubled target space}
\label{AKSZdouble}

Let us recall a key point in the analysis of Ref. \cite{Mylonas:2012pg}. Consider the sigma model 
(\ref{aksz}) associated to the standard CA on a torus. Moreover let $T=R$ be the only generalized 
3-form with $R$ a constant 3-vector. This means that the 3D action is{\footnote{As mentioned in 
Ref. \cite{Mylonas:2012pg} a 2D kinetic term should be included for consistency with the equations 
of motion. We do not explicitly write it here because it will not play a crucial role in the argument.}}
\be 
S_{R}[X,A,F]=\int_{\S_3}\biggl(F_a\w\dd X^a+q^a\w\dd p_a-q^a\w F_a+\sfrac 16R^{abc}p_a\w p_b\w p_c\biggl)~,
\ee
where we used only early Latin indices because for the moment we refer to the flat torus.
Integrating out the 2-form $F_a$ one obtains 
\be 
S_{R}[X,A,F]=\int_{\S_2}p_a\w\dd X^a+\int_{\S_3}\sfrac 16 R^{abc}p_a\w p_b\w p_c~,
\ee
with $X^a$-independent $R^{abc}$ by assumption. The equation of motion for $X^a$ is simply
\be 
\dd p_a=0~.
\ee
This means that the 1-form $p_a$ may be written locally as 
\be 
p_a=\dd {\tilde X}_a~,
\ee
where $\tilde X_a\in C^{\infty}(\S_3,X^{\star}\text{T}^{\star}\text{M})$. These $\tilde X_a$ are similar 
to the dual coordinates of DFT, which is the reason for our choice of notation. 
As suggested in Ref. \cite{Mylonas:2012pg}, in the sigma model they essentially correspond to an augmented embedding of the 2-dimensional boundary theory on $\S_2$ 
in the 
full cotangent bundle of the target manifold M. In other words there are generalized (or doubled) target space coordinates 
$(\X^I)=(X^a,\tilde X_a)$ which correspond to the map $\X=(\X^I):\S_3\to \text{T}^{\star} {\text{M}}$. Note that the appearance of the dual coordinates is very natural in 
this context, since they were suggested by the equations of motion of the sigma model.

An alternative way to think about the above doubling is in the spirit of the topological approach to T-duality 
\cite{CAred1,CAred2}, which was explained via Courant algebroids in Refs. \cite{CAred3,CAred4,CAred5}. 
In this approach there is a product manifold $\text{M}\times \tilde{\text{M}}$ of original and dual spaces and 
T-duality corresponds to an isomorphism of twisted K-theories \cite{CAred1,CAred2}. In \cite{CAred4} 
it was shown that this can be extended to an isomorphism between the corresponding CAs. 
Here we associate $\X^I$ to the product manifold $\text{M}\times \tilde{\text{M}}$. Presumably, the AKSZ 
sigma models for CAs over this extended
target space correspond to the ones we will consider below. We plan to study this correspondence carefully
 in future work.

Once one considers the possibility of such generalized embeddings, it is natural to allow all the fields that 
appear in the model to depend both on $X^a$ and $\tilde X_a$. In that case
insisting on the formulation (\ref{aksz}) for the sigma model is rather restrictive. From the viewpoint of physics, 
Eq. (\ref{aksz}) does not contain $d\tilde X_a$ at all,
which should not be the case in general.
Thus, returning to the general case, our proposal here is twofold. First, allow $B,\beta,h, a$ and $T$ to depend on both $X^a$ and $\tilde X_a$. Second, 
introduce a second world volume 2-form $\tilde F^a \in \Omega^2(\S_3,\X^{\star}\text{TM})$; note that this is again an 
 auxiliary world volume 2-form like $F_a$, with the difference of having a \emph{vector} index  instead.  
Then we write the 3-dimensional action
\be \label{akszDouble}
S_{\S_3}=\mkern-3mu\int_{\S_3}\mkern-3mu\biggl(F_{a}\w\dd X^{a}+\tilde F^a\w\dd \tilde{X}_a+\sfrac 12 \eta_{IJ}A^I\w\dd A^J-P_{I}^aA^I\w F_a-
\tilde P_{aI}A^I\w\tilde F^a+\sfrac 16T_{IJK}A^I\w A^J\w A^K\biggl)~.
\ee
In more compact notation, writing  $P_I^J=(P^a_I,\tilde P_{aI})$ and  $F^I=(F_a,\tilde F^a)$ for $F^I\in\Omega^2(\S_3,\X^{\star} E)$, we get
\be 
S_{\S_3}=\int_{\S_3}\biggl(\d_{IJ}F^{I}\w\dd \X^{J}+\sfrac 12 \eta_{IJ}A^I\w\dd A^J-\d_{JK}P_{I}^JA^I\w F^K
+\sfrac 16T_{IJK}A^I\w A^J\w A^K\biggl)~.
\ee
The boundary action is the same as before, namely
\be 
S_{\S_2}=\int_{\S_2}\sfrac 12 {\cal B}_{IJ}A^I\w A^J~,
\ee
with the difference that ${\cal B}={\cal B}(X,\tilde X)$. An important remark regards the object $\tilde P_{aI}$, which was 
absent before. These are the components of a map $\tilde P:E\to \ctb$ that maps elements of the Courant algebroid to 
the cotangent bundle. Examples of such a map is the unit map on 1-forms and the map $B^{\sharp}: \tb \to \ctb$ that acts simply as 
$B^{\sharp}(X_i)=B_{ij}\eta^j$.

Our purpose now is to consider the analog of the construction we did for the CA $L_{B\beta}\oplus L_{B\beta}^{\star}$, bearing 
in mind that a complete mathematical characterization of the construction is due. 
The ingredients are similar to the standard case. We consider the twists $\phi$ and $\psi$, as given in 
Eqs. \eqref{phiexp2} and \eqref{psiexp2}, as well as the geometric 
twist $f$ of the nilmanifold that appears in Eq. \eqref{f}. Then the action reads as 
\bea 
S&=&\int_{\S_3}\biggl(F_{a}\w\dd X^{a}+\tilde F^a\w\dd \tilde{X}_a+kq^i\w\dd p_i+k' p_i\w\dd q^i\nn\\
&&-(\mu e^a_iq^i+\nu\beta^{ij}e_j^ap_i)
\w F_a-(\mu' e_a^ip_i+\nu' B_{ij}e^j_aq^i)
\w \tilde F^a+f-\phi-\psi\biggl)\nn\\
&&+\int_{\S_2}\biggl(\sfrac 12 B_{ij}q^i\w q^j+\sfrac 12 \beta^{ij}p_i\w p_j+\sfrac 12 h_i^jq^i\w p_j\biggl)~.\label{akszdoubledexplicit}
\eea 
For the map $\tilde P$ we took 
\be
\tilde P_a^i=\mu' e^i_a \quad \text{and}\quad \tilde P_{ai}=\nu' B_{ij}e^j_a~,
\ee
which is the natural choice. As before, the parameters $\mu,\nu,\mu',\nu'$ are valued in $\{0,1\}$, which reflects the 
flexibility of trivializing the corresponding map or not. Once more, $k$ and $k'$ should satisfy $k+k'=1$. 

Next we determine the equations of motion on the boundary by varying with respect to $X^a, \tilde X_a, q^i$ and $p_i$. 
The only new equation is 
\be 
\d_{\tilde X_a}S|_{\S_2}=\tilde F^a+\sfrac 12 \tilde\partial^aB_{jk}q^j\w q^k+\sfrac 12 \tilde\partial^a\beta^{jk}p_j\w p_k+\sfrac 12 \tilde \partial^ah
_j^kq^j\w p_k=0~,
\ee
where $\tilde\partial^a=\partial/\partial\tilde X_a$. The other three equations are exactly as in \eqref{eoms}. 
Additionally, the bulk/boundary condition that should hold reads as 
\be\label{bbdouble}
(\mu e^a_iq^i+\nu\beta^{ij}e_j^ap_i)
\w F_a+(\mu' e_a^ip_i+\nu' B_{ij}e^j_aq^i)
\w \tilde F^a=f-\phi-\psi \qquad \text{on $\S_2$}~.
\ee
This has to be consistent with the choice of boundary conditions that guarantee the equations of motion on the boundary. 

Let us examine how the boundary conditions that were considered in section \ref{BbetaAKSZ} are modified. First we consider 
the boundary conditions
\bea 
F_a|_{\S_2}&=&-\sfrac 12\partial_a{B}_{jk}q^j\w q^k-\sfrac 12 \partial_a{\beta}^{jk}p_j\w p_k-\sfrac 12\partial_a{h}_j^kq^j\w p_k~,
\nn\\
\tilde F^a|_{\S_2}&=&-\sfrac 12 \tilde\partial^aB_{jk}q^j\w q^k-\sfrac 12 \tilde\partial^a\beta^{jk}p_j\w p_k-\sfrac 12 \tilde \partial^ah
_j^kq^j\w p_k~,\nn\\
\d q^i|_{\S_2}&=&0~,\nn\\
(kq^i+{\beta}^{ij}p_j-\sfrac 12{h}^i_jq^j)|_{\S_2}&=&0~.\label{bcdouble1}
\eea
The bulk/boundary consistency condition \eqref{bbdouble} becomes formally identical to \eqref{condition1}, namely 
\begin{align}
{\cal R}^{ijk}
-\sfrac 1k {\cal Q}_n^{[ij}\beta^{\underline{p}k]}\chi^{n}_p
+\sfrac 1{k^2}{\cal F}^{[i}_{mn}\beta^{\underline{p}j}\beta^{\underline{q}k]}
\chi^{m}_p\chi^{n}_q
-\sfrac 1{k^3}{\cal H}_{lmn} \beta^{pi}\beta^{qj}\beta^{rk}
\chi^{l}_p\chi^{m}_q\chi^{n}_r=0~,
\nn
\end{align}
but with the 
upgraded definitions
\bea 
&&\mkern-30mu{\cal R}^{ijk}=\psi^{ijk}-3\nu\beta^{[i\underline{l}}\theta_l{\beta}^{jk]}+\beta^{li}\beta^{mj}\beta^{nk}\phi_{lmn}-3\mu'\tilde\theta^{[i}\beta^{jk]}~,
\nn\\
&&\mkern-30mu{\cal Q}^{ij}_{k}=-3\mu \theta_k{\beta}^{ij}+3\nu\beta^{[i\underline{l}}\theta_l{h}^{j]}_k+3B_{lk}\psi^{ijl}
+3(1+\beta B)^l_k\beta^{mi}\beta^{nj}\phi_{lmn}+3\mu'\tilde\theta^{[i}h^{j]}_k-3\nu'B_{kl}\tilde\theta^l\beta^{ij},\nn\\
&&\mkern-30mu{\cal F}^i_{jk}=-3\mu\theta_{[j}{h}_{k]}^{i}-3f^{i}_{jk}-3\nu\beta^{il}\theta_l{B}_{jk}+3B_{lj}B_{mk}\psi^{lmi}
+3(1+\beta B)^l_j(1+\beta B)^m_k\beta^{ni}\phi_{lmn}\nn\\
&&\quad-3\mu'\tilde\theta^{i}B_{jk}-3\nu'B_{[j\underline{l}}\tilde\theta^l h^{i}_{k]}~,\nn\\
&&\mkern-30mu{\cal H}_{ijk}= (1+\beta B)^l_i(1+\beta B)^m_j(1+\beta B)^n_k\phi_{lmn}-3\mu\theta_{[i}{ B}_{jk]}
+B_{li}B_{mj}B_{nk}\psi^{lmn}-3\nu'B_{[i\underline{l}}\tilde\theta^{l}B_{jk]}~,\nn\\\label{hfqrdouble}
\eea
where we defined $\tilde\theta^i=e^i_a\tilde\partial^a$. 
Similarly, the boundary conditions 
\bea 
F_a|_{\S_2}&=&\sfrac 12\partial_a{B}_{jk}q^j\w q^k+\sfrac 12 \partial_a{\beta}^{jk}p_j\w p_k+\sfrac 12\partial_a{h}_j^kq^j\w p_k~,
\nn\\
\tilde F^a|_{\S_2}&=&\sfrac 12 \tilde\partial^aB_{jk}q^j\w q^k+\sfrac 12 \tilde\partial^a\beta^{jk}p_j\w p_k+\sfrac 12 \tilde \partial^ah
_j^kq^j\w p_k~,\nn\\
(k'p_i-{B}_{ij}q^j-\sfrac 12{h}_i^jp_j)|_{\S_2}&=&0~,\nn\\
\d p_i|_{\S_2}&=&0~,\label{bcdouble2}
\eea
lead to the generalization of the alternative condition \eqref{condition2}, namely
\begin{align}
 {\cal H}_{ijk}-\sfrac 1{k'}{\cal F}_{[ij}^n
{B}_{\underline{p}k]}\chi'^p_n
+\sfrac 1{k'^2}{\cal Q}_{[i}^{mn}{B}_{\underline{p}j}B_{\underline{q}k]}\chi'^p_m\chi'^q_n
-\sfrac 1{k'^3}{\cal R}^{lmn}
{B}_{pi}{B}_{qj}{B}_{rk}\chi'^p_l\chi'^q_m\chi'^r_n=0~,\nn
\end{align}
with the definitions \eqref{hfqrdouble}.

\subsection{The pure $R$ flux limit}
\label{pureR}

Let us briefly revisit the pure $R$ flux limit of Ref. \cite{Mylonas:2012pg} with the results of Section \ref{AKSZdouble}. 
Consider $B=h=0$ and $\beta=\beta(\tilde X)$ to be independent of $X^a$. Additionally, let us turn off the $\phi$ flux and geometric flux 
$f$ just for the present example, namely $f=\phi=0$. With these assumptions, Eqs. (\ref{hfqrdouble}) reduce to
\bea 
&& {\cal R}^{ijk}=\psi^{ijk}-3\tilde\partial^{[i}\beta^{jk]}~,\nn\\
&& {\cal Q}={\cal F}={\cal H}=0~.
\eea
We choose the boundary conditions (\ref{bcdouble1}), in which case the bulk/boundary condition (\ref{condition1}) reads as
\be 
{\cal R}^{ijk}=0\quad \Rightarrow \quad \psi^{ijk}=3\tilde\partial^{[i}\beta^{jk]}~.
\ee
In case $\beta$ is linear in $\tilde X_a$, e.g. $\beta=\tilde N\epsilon^{ijk}\d_k^a\tilde X_ap_i\w p_j$, we can identify $\psi$ with the constant $R$ flux, 
e.g. $R=\tilde N$. This is similar to the case considered in 
\cite{Mylonas:2012pg}, where more details may be found.

\subsection{Genuine non-geometry?}

The main motivation for the formulation we propose in Section \ref{AKSZdouble} was to examine the possibility to construct 
genuinely non-geometric 
models, in the sense that was explained in the Introduction. This is essentially the message of the 
results in Section \ref{AKSZdouble}, but in order to make sure that they are not empty and indeed contain nontrivial cases 
we construct here an explicit toy model. 

Consider the sigma model on the twisted torus of Section \ref{akszexplicit}, upgraded to 
a model of the type (\ref{akszdoubledexplicit}) with the following background fields instead:
\be 
B=NX^1q^2\w q^3~,\quad \beta=\tilde N\tilde X_2p_1\w p_3~.
\ee
Both $B$ and $\beta$ are nonvanishing, and they satisfy
\be
\beta B=\begin{pmatrix}
         0 & -N\tilde NX^1\tilde X_2 & 0  \\ 0 & 0 & 0 
          \\ 0 & 0 & 0
        \end{pmatrix}~.
\ee
Moreover, in this case we identify $\phi$ and $\psi$ with the corresponding derivations, namely
\be 
\phi=Nq^1\w q^2\w q^3~,\quad \psi=\tilde Np_1\w p_2\w p_3~.
\ee 
Making the choices $k=k'=\sfrac 12$ and $\mu=\mu'=\nu=\nu'=1$~, the model is given as
\bea  
S=\int_{\S_3}&&\biggl(F_a\w\dd X^a+\tilde F^a\w\dd \tilde X_a+\sfrac 12 p_i\w\dd q^i+\sfrac 12 q^i\w\dd p_i\nn\\
&&-F_1\w(q^1-\tilde N\tilde X_2 p_3)-F_2\w q^2-F_3\w (q^3+X^1q^2+\tilde N\tilde X_2p_1)\nn\\
&&-\tilde F^1\w p_1-\tilde F^2\w(p_2-X^1p_3-NX^1q^3-N(X^1)^2q^2)-\tilde F^3\w(p_3+NX^1q^2)\nn\\
&&+q^1\w q^2\w p_3-\phi-\psi \biggl)\nn\\
+\int_{\S_2}&&\biggl(NX^1q^2\w q^3+\tilde N\tilde X_2p_1\w p_3\biggl)~,
\eea
where in the present case we find
\begin{flalign}
&&\phi= Nq^1\w q^2\w q^3\mkern-3mu+\mkern-3muN\tilde N\tilde X_2(q^2\w q^3\w p_3+q^2\w q^1\w p_1)\mkern-3mu+\mkern-3mu N(\tilde N\tilde X_2)^2q^2\w p_3\w p_1~,\nn\\
&&\psi =\tilde Np_1\w p_2\w p_3\mkern-3mu+\mkern-3muN\tilde N X^1(p_3\w p_1\w q^3\mkern-2mu+\mkern-2mu p_2\w p_1\w q^2)\mkern-3mu+\mkern-3mu\tilde N(NX^1)^2p_1\w q^2\w q^3~.
\end{flalign}
The equations of motion for $X^a$ and $\tilde X_a$ lead to the boundary conditions for $F_a$ and $\tilde F^a$:
\bea 
&& F_1=-Nq^2\w q^3~,\quad F_2=F_3=0~,\nn\\
&& \tilde F^2=-\tilde Np_1\w p_3~,\quad \tilde F^1=\tilde F^3=0~.
\eea
Additionally, the equations of motion for $q^i$ and $p_i$ lead to:
\bea
&& (\sfrac 12 q^1+\tilde N\tilde X_2p_3)\d p_1=0~,\quad  (\sfrac 12 q^2)\d p_2=0~,\quad  (\sfrac 12 q^3-\tilde N\tilde X_2)\d p_3=0~,\nn\\ 
&& (\sfrac 12p_1)\d q^1=0~, \quad (\sfrac 12 p_2+NX^1q^3)\d q^2=0~, \quad  (\sfrac 12 p_3-NX^1q^2)\d q^3=0~.\label{varsexample2}
 \eea
 An analysis similar to Section \ref{akszexplicit} dictates the boundary conditions
 \be 
 \d q^1=0~,\quad \d p_2=0~,\quad \big(\d p_3=0 \quad \text{or}\quad \d q^2=0\big)~, 
\quad \big(\d p_1=0 \quad \text{or}\quad \d q^3=0\big)~. 
 \ee
 Out of the last two requirements we make the random choice $\d p_3=\d q^3=0$.
 The remaining boundary conditions are
 \be 
 q^1=-2\tilde N\tilde X_2p_3~,\quad p_2=-2NX^1q^3~.
 \ee
 Imposing also that $X^1p_1-2\tilde X_2q^2=0$~, we find that on the boundary
 \bea
\phi&=&-N\tilde N\tilde X_2p_3\w q^2\w q^3~,\nn\\
\psi&=&N\tilde NX^1p_1\w p_3\w q^3~,
 \eea
 and that the bulk/boundary consistency condition
 \be 
 F_1\w(q^1-\tilde N\tilde X_2 p_3)
+\tilde F^2\w(p_2-X^1p_3-NX^1q^3-N(X^1)^2q^2)
+q^1\w q^2\w p_3-\phi-\psi=0~,
 \ee
is satisfied. This means that the model is a nontrivial case where the twists $\phi$ and $\psi$, as well as the deformations $B$ and 
$\beta$, are nonvanishing. Unlike the model with pure 3-vector flux, which 
is well known to be T-dual to standard geometric models, the present case cannot be T-dualized to a standard geometry. 
Thus it constitutes a genuine case of non-geometry. 

The latter statement is corroborated by attempting to write down the corresponding 2D string model. 
This is not possible just in terms of $X^a$; instead $\tilde X_a$ necessarily appear, similarly to the pure 
$R$-flux models considered in Refs. \cite{Halmagyi:2009te,Mylonas:2012pg} but in a significantly more complicated way.
The topological sector of the corresponding   model 
can be written as 
\bea 
&&\mkern-45mu\int_{\S_3}\biggl(8N^2\tilde N\tilde X_1\e^1\w e^2\w e^3+4N\tilde N^2X^2\tilde X_2\dd\tilde X_1\w\dd\tilde X_2\w\dd\tilde X_3 +N\tilde N\tilde X_2e^1\w e^2\w\dd\tilde X_1+\nn\\
&&\mkern-45mu-2N\tilde N(\tilde X_2 e^2-\tilde X_1 e^1)\w e^3\w\dd\tilde X_3\biggl)+\nn\\
&&\mkern-45mu +\int_{\S_2} \biggl(-NX^1(1+4N\tilde NX^1\tilde X_1)e^2\w e^3+\tilde N(\tilde X_2+N X^2(X^1)^2+2 N\tilde N X^2 \tilde X_2^2)\dd \tilde X_1\w\dd\tilde X_3
+\nn\\
&&\mkern-45mu+\frac{3}{2}N\tilde NX^1\tilde X_2e^2\w\dd\tilde X_1-2N\tilde NX^1\tilde X_1(e^3-X^2e^1)\w\dd\tilde X_3\biggl)~,
\eea
which supports the above remarks. Notice that for $\tilde N\to 0$, namely when the 
deformation $\beta$ is turned off, we obtain 
\be\int_{\S_2} -NX^1e^2\w e^3~,\ee
while for $N\to 0$ (or $B=0$) we obtain 
\be\int_{\S_2}\tilde N\tilde X_2\dd \tilde X_1\w\dd\tilde X_3~,\ee
as expected. 

\section{Conclusions}

The extended nature of the fundamental degrees of freedom in string theory leads to duality symmetries, whose consequences are 
unconventional from a traditional field theory viewpoint. One of these consequences is our encounter with non-geometric 
string backgrounds. A central question in this line of research is whether such backgrounds are always equivalent (up to 
duality) to previously known geometric ones or there exist ones that are truly new. Recent developments, mainly in the context of 
DFT, suggest that duality orbits of flux configurations that do not intersect geometric regions indeed exist \cite{Dibitetto:2012rk}.

In this paper we addressed the problem of constructing sigma models that correspond to genuinely non-geometric backgrounds. 
This approach is inspired by previous work along these lines in the string theory literature \cite{Halmagyi:2009te,Mylonas:2012pg}, 
which we extended and generalized. The underlying mathematical setting is that of Courant algebroids, which has recently found 
applications in the physics of string theory 
\cite{Halmagyi:2009te,Blumenhagen:2012pc,Mylonas:2012pg,Blumenhagen:2013aia,Chatzistavrakidis:2013wra}. Here we constructed 
a general class of CAs with base manifolds being twisted tori. The choice of twisted tori is made for a number of reasons, 
in particular (i) they are the simplest nontrivial generalization of flat tori that retain parallelizability and they can be endowed 
with all kinds of generalized complex structures \cite{sixmanifolds}{\footnote{For example, 
several twisted tori admit 
a symplectic structure and their phase space can be completely characterized, see \cite{Chatzistavrakidis:2014tda}.}}, 
(ii) they naturally incorporate geometric fluxes, and (iii) they play a central role in flux compactifications, notably in 
Scherk-Schwarz reductions. We followed the approach of introducing the basic mathematical notions first,
 then applying 
them for general twisted tori of step 2, and finally examining in detail an illustrative example from the class.

In order to reach our main goal of constructing relevant sigma models, we resided on the result that every CA structure over a manifold 
M has an associated topological sigma model with M as target space \cite{Roytenberg:2006qz}. For physical applications, it is 
natural to consider manifolds with boundary and add general topological boundary terms and also kinetic terms that break the topological 
nature of the model. Studying the corresponding membrane sigma models for the class of CAs we constructed, we found general 
bulk/boundary consistency conditions appearing in Eqs. \eqref{condition1} and \eqref{condition2}. These expressions generalize 
on one hand previously known integrability conditions for Dirac structures \cite{severa,sw,Hofman:2002rv}, and on the other hand allow 
for a systematic characterization of fluxes, extending expressions found in \cite{Halmagyi:2009te}. In certain limits, our expressions 
reproduce previous results; 
on the other hand we also studied in detail a case where both 2-form and 2-vector deformations 
coexist meaningfully without being inverse of one another. 

However, in order to really account for cases that appear in string theory via generalized T-duality, the above sigma models 
cannot be the end of the story. This was already noticed in \cite{Halmagyi:2009te}, and later in \cite{Mylonas:2012pg}, where 
sigma models of an extended type were first suggested. These sigma models have the phase space of M as target space, instead of M itself. 
Inspired by this approach, we proposed a minimal systematic generalization of the previous sigma models that incorporates this doubled 
point of view. Analysing such models we found that the bulk/boundary consistency conditions take again the form 
appearing in Eqs. \eqref{condition1} and \eqref{condition2}, albeit with an upgraded set of definitions that characterize the 
fluxes of the model. Then we were able to write down an explicit example of a model which combines the following properties:  
(i) all types of generalized fluxes are present, (ii) it cannot be reduced to a 2D theory with standard target space and 
(iii) it cannot be dualized to a standard geometric model. This makes this example, and any other constructed similarly, an 
excellent toy model for genuinely non-geometric backgrounds.

Finally, it is interesting to compare our results with the expressions for fluxes found in the 
context of DFT and its generalized Scherk-Schwarz dimensional reduction on twisted doubled tori 
\cite{dftflux1,dftflux2,dftflux3,dftflux4}. For this comparison, it is not enough to look at Eqs. \eqref{hfqrdouble}, 
which contain less information than the corresponding ones from DFT. The relevant equations are instead the full conditions 
\eqref{condition1} and \eqref{condition2} with the definitions \eqref{hfqrdouble}. These two equations give the ``$H$'' and 
``$R$'' flux in the present formulation, which actually contain all terms appearing in DFT plus additional terms of higher order in the 
combinations of $B$ and $\beta$. For the other two sets of fluxes the comparison is not yet possible, since we have not determined general 
expressions for mixed boundary conditions in this paper. However, we can conclude that our formulation encompasses results from DFT 
and it would be interesting to examine further the relation between DFT (target space theory) and the sigma model we proposed.

\paragraph{Acknowledgements.} The authors thank D.~Berman, 
F.~F.~Gautason,
D.~Roytenberg, P.~Schupp and P.~\v{S}evera for 
discussions. A.C. and L.J. have greatly benefited from discussions and collaboration with A.~Deser and 
T.~Strobl in the framework of a related project. 
The work of L.J. was supported in part  by the  Alexander von Humboldt Foundation and by Croatian Science Foundation under the project IP-2014-09-3258.

\appendix 

\section{Proof of protobialgebroid structure on $(L_{B\beta},L^{\star}_{B\beta})$}

In this appendix we prove that $(L_{B\beta},L^{\star}_{B\beta})$ with the ingredients (brackets, anchors and twists) 
given in Section \ref{pbanil} is a protobialgebroid, i.e. it satisfies the properties 1 to 4 of Definition \ref{pba}.

\paragraph{Proof of property 1.} For $X=X^i\Theta_i$ and $Y=Y^i\Theta_i$ we compute:
\bea
[X,fY]_{L_{B\beta}}&=&e^Be^{\beta}\biggl([e^{-\beta}e^{-B}X,e^{-\beta}e^{-B}fY]_{\text{Lie}}+\beta(e^{-\beta}e^{-B}(\phi(X,fY,\cdot)),\cdot)\biggl)
\nn\\
&=&e^{B}e^{\beta}\biggl([e^{-\beta}e^{-B}X,fe^{-\beta}e^{-B}Y]_{\text{Lie}}+f\beta(e^{-\beta}e^{-B}(\phi(X,Y,\cdot)),\cdot)\biggl)\nn\\
&=&e^Be^{\beta}\biggl(f[e^{-\beta}e^{-B}X,e^{-\beta}e^{-B}Y]_{\text{Lie}}+(e^{-\beta}e^{-B}X(f))e^{-\beta}e^{-B}Y\nn\\
&&+f\beta(e^{-\beta}e^{-B}(\phi(X,Y,\cdot)),\cdot)\biggl)
\nn\\
&=&f\biggl(e^Be^{\beta}\big([e^{-\beta}e^{-B}X,e^{-\beta}e^{-B}Y]_{\text{Lie}}+\beta(e^{-\beta}e^{-B}(\phi(X,Y,\cdot)),\cdot)\big)\biggl)
\nn\\
&&+e^Be^{\beta}(\rho(X)f)e^{-\beta}e^{-B}Y\nn\\
&=&f[X,Y]_{L_{B\beta}}+(\rho(X)f)Y~.
\eea

Similarly the proof for $\eta=\eta_iE^i$ and $\xi=\xi_iE^i$.

\paragraph{Proof of property 2.} 

Essentially this is already covered by construction in the main text. 
As a cross check we compute:
\bea
\rho([X,Y]_{L_{B\beta}})&=&\rho\biggl(e^Be^{\beta}\big([e^{-\beta}e^{-B}X,e^{-\beta}e^{-B}Y]_{\text{Lie}}+
\beta(e^{-\beta}e^{-B}(\phi(X,Y,\cdot)),\cdot)\big)\biggl) \nn\\
&\overset{\eqref{anchor1},\eqref{anchor2}}=&\rho\biggl(e^Be^{\beta}\big([\rho(X),\rho(Y)]_{\text{Lie}}+
\rho_{\star}(\phi(X,Y,\cdot))\big)\biggl) \nn\\
&\overset{\eqref{anchor1}}=&[\rho(X),\rho(Y)]_{\text{Lie}}+
\rho_{\star}(\phi(X,Y,\cdot))
\eea
and the proof is complete. Similarly for the corresponding property on $L^{\star}_{B\beta}$.

\paragraph{Proof of property 3.}

We directly apply the general expressions for the derivations on $L_{B\beta}$ and $L^{\star}_{B\beta}$:
\bea
\dd_{L_{B\beta}} \o(X_1,\dots ,X_{p+1})&=&\sum_{i=1}^{p+1}(-1)^{i+1}
\rho(X_i)\o(X_1,\dots,\hat X_i,\dots ,X_{p+1})+\nn\\ 
&+&\sum_{i<j}(-1)^{i+j}\o([X_i,X_j]_{L_{B\beta}},X_1,\dots ,\hat X_i,\dots ,\hat X_j, \dots ,X_{p+1})~.\nn\\
\dd_{L_{B\beta}^{\star}} \Omega(\eta_1,\dots ,\eta_{p+1})&=&\sum_{i=1}^{p+1}(-1)^{i+1}
\rho_{\star}(\eta_i)\Omega(\eta_1,\dots,\hat \eta_i,\dots ,\eta_{p+1})+\nn\\ 
&+&\sum_{i<j}(-1)^{i+j}\Omega([\eta_i,\eta_j]_{L_{B\beta}^{\star}},\eta_1,\dots ,\hat \eta_i,\dots ,\hat \eta_j, \dots ,\eta_{p+1})~,\nn
\eea
for arbitrary $\o\in\G(\w^p L_{B\beta}^{\star})$ and $\Omega\in \G(\w^pL_{B\beta})$, to compute the derivations of the 
basis elements $E^i\in \G(\w^1L^{\star}_{B\beta})$ and $\Theta_i\in\G(\w^1L_{B\beta})$
\bea 
\dd_{L_{B\beta}}E^i&=&-\sfrac 12 (f^i_{jk}-\beta^{il}\phi_{ljk})E^j\w E^k~,\label{dE}\\
\dd_{L^{\star}_{B\beta}}\Theta_i&=&-\sfrac 12 (\theta_i\beta^{jk}+2\beta^{jm}f^k_{im}+\beta^{jl}\beta^{km}\phi_{ilm})\Theta_j\w\Theta_k~.
\label{dTheta}
\eea
Then we compute
\be
[[\Theta_i,\Theta_j]_{L_{B\beta}},\Theta_k]_{L_{B\beta}}=\big(\theta_k(\beta^{mn}\phi_{mij})-f_{ij}^l\beta^{nm}\phi_{mlk}-f^n_{lk}\beta^{lm}\phi_{mij}+
\beta^{lm}\beta^{np}\phi_{mij}\phi_{pkl}\big)\Theta_n~,
\ee
and 
\be 
\phi(\dd_{L^{\star}_{B\beta}}\Theta_i,\Theta_j,\Theta_k)=-\phi_{jkl}\big(\theta_i\beta^{ln}-2f_{im}^{[l}\beta^{n]m}+\beta^{lp}\beta^{nm}\phi_{ipm}\big)\Theta_n~.
\ee
Moreover,
\be
\dd_{L^{\star}_{B\beta}}\phi(\Theta_i,\Theta_j,\Theta_k)=\beta^{lm}\theta_m\phi_{ijk}\Theta_n~.
\ee
These expressions deliver the result
\bea
&& \mkern-45mu[[\Theta_i,\Theta_j]_{L_{B\beta}},\Theta_k]_{L_{B\beta}}+\text{c.p.}-\dd_{L^{\star}_{B\beta}}\phi(\Theta_i,\Theta_j,\Theta_k)-
\phi(\dd_{L^{\star}_{B\beta}}\Theta_i,\Theta_j,\Theta_k)-\phi(\Theta_i,\dd_{L^{\star}_{B\beta}}\Theta_j,\Theta_k)\nn\\
&& \mkern-45mu - \phi(\Theta_i,\Theta_j,\dd_{L^{\star}_{B\beta}}\Theta_k)=\beta^{ml}\big(\theta_{[i}\phi_{jkl]}-\sfrac 32f^n_{[ij}\phi_{kl]n}+\sfrac 32
\beta^{np}\phi_{n[ij}\phi_{kl]p}\big)\Theta_m~,
\eea
which means that the property holds when the condition
\be \label{bi1}
\theta_{[i}\phi_{jkl]}-\sfrac 32\phi_{m[ij}(f^m_{kl]}-\beta^{nm}\phi_{\underline{n}kl]})=0
\ee
is satisfied. A similar computation for the dual property yields the condition
\be \label{bi2}
\beta^{[l\underline{m}}\theta_m\psi^{ijk]}-\sfrac 32 \psi^{m[jk}(\theta_m\beta^{li]}+\beta^{l\underline{n}}f^{i]}_{mn}+
\beta^{l\underline{s}}\beta^{i]t}\phi_{mst})=0~.
\ee
These two conditions are essentially Bianchi identities as will be clear from property 4.

\paragraph{Proof of property 4.}
Using the expansions 
\bea
\phi&=&\sfrac 16\phi_{ijk}E^{i}\w E^j\w E^k~,\nn\\
\psi&=&\sfrac 16 \psi^{ijk}\Theta_i\w\Theta_j\w\Theta_k~,
\eea 
and the result \eqref{dE} 
we compute
\bea 
\dd_{L_{B\beta}}\phi&=&\sfrac 16(\dd_{L_{B\beta}}\phi_{ijk})E^i\w E^j\w E^k+\sfrac 12 \phi_{ijk}(\dd_{L_{B\beta}}E^i)\w E^j\w E^k 
  \nn\\
&=&\sfrac 16\big( \theta_l \phi_{ijk}-\sfrac 32\phi_{mjk}(f^m_{li}-\beta^{mn}\phi_{nli})\big)E^{l}\w E^i\w E^j\w E^k~,
\eea
which vanishes when the condition \eqref{bi1} is satisfied. This is essentially a Bianchi identity 
(and fully agrees with previous results, e.g.
\cite{Blumenhagen:2012pc,dftflux4}).
It is simple to check that this Bianchi identity is satisfied in the example of section \ref{example}.
On the other hand, using \eqref{dTheta} we compute 
\bea 
\mkern-30mu\dd_{L_{B\beta}^{\star}}\psi&=&\sfrac 16 (\dd_{L_{B\beta}^{\star}}\psi^{ijk})\Theta_i\w\Theta_j\w\Theta_k+\sfrac 12 \psi^{ijk}
(\dd_{L^{\star}_{B\beta}}\Theta_i)\w\Theta_j\w \Theta_k\nn\\
&=&\sfrac 16(\beta^{lm}\theta_m\psi^{ijk}-\sfrac 32 \psi^{mjk}(\theta_m\beta^{li}+\beta^{ln}f^i_{mn}+\beta^{ls}\beta^{it}\phi_{mst})
\Theta_l\w\Theta_i\w\Theta_j\w \Theta_k~,
\eea
which vanishes when the Bianchi identity \eqref{bi2} is satisfied.
This is also true in the example of section \ref{example}.


\begin{thebibliography}{99}
 \addtolength{\itemsep}{-3pt}
  
\bibitem{Shelton:2005cf}
  J.~Shelton, W.~Taylor and B.~Wecht,
  JHEP {\bf 0510} (2005) 085
  [hep-th/0508133].
  
\bibitem{Aldazabal:2006up}
  G.~Aldazabal, P.~G.~Camara, A.~Font and L.~E.~Ibanez,
  JHEP {\bf 0605} (2006) 070 \\  
{}  [hep-th/0602089].
  
\bibitem{Halmagyi:2008dr}
  N.~Halmagyi,
  JHEP {\bf 0807} (2008) 137
  [arXiv:0805.4571 [hep-th]].
  
\bibitem{Halmagyi:2009te}
  N.~Halmagyi,
  ``Non-geometric Backgrounds and the First Order String Sigma Model,''
  arXiv:0906.2891 [hep-th].
 
\bibitem{Blumenhagen:2012pc}
  R.~Blumenhagen, A.~Deser, E.~Plauschinn and F.~Rennecke,\\
Class.\ Quant.\ Grav.\  {\bf 29} (2012) 135004
  [arXiv:1202.4934 [hep-th]].
 
\bibitem{Mylonas:2012pg}
  D.~Mylonas, P.~Schupp and R.~J.~Szabo,
  JHEP {\bf 1209} (2012) 012 \\  
{}  [arXiv:1207.0926 [hep-th]].

\bibitem{Blumenhagen:2013aia}
  R.~Blumenhagen, A.~Deser, E.~Plauschinn, F.~Rennecke and C.~Schmid, \\
  Fortsch.\ Phys.\  {\bf 61} (2013) 893
  [arXiv:1304.2784 [hep-th]].

  
\bibitem{Chatzistavrakidis:2013wra}
  A.~Chatzistavrakidis, L.~Jonke and O.~Lechtenfeld,
  Nucl.\ Phys.\ B {\bf 883} (2014) 59 \\ {}
  [arXiv:1311.4878 [hep-th]].
  
  \bibitem{gcg2}
  M.~Gualtieri,
  ``Generalized complex geometry,'' 
  DPhil thesis, \\
math/0401221 [math.DG].
  
  
\bibitem{Hull:2004in}
  C.~M.~Hull,
  JHEP {\bf 0510} (2005) 065
  [hep-th/0406102].

\bibitem{Dabholkar:2005ve}
  A.~Dabholkar and C.~Hull,
  JHEP {\bf 0605} (2006) 009
  [hep-th/0512005].
  
\bibitem{Hull:2007jy}
  C.~M.~Hull and R.~A.~Reid-Edwards,
  JHEP {\bf 0808} (2008) 043
  [arXiv:0711.4818 [hep-th]].
  
\bibitem{Hull:2009sg}
  C.~M.~Hull and R.~A.~Reid-Edwards,
  JHEP {\bf 0909} (2009) 014
  [arXiv:0902.4032 [hep-th]].
  
\bibitem{Narain:1986qm}
  K.~S.~Narain, M.~H.~Sarmadi and C.~Vafa,
  Nucl.\ Phys.\ B {\bf 288} (1987) 551.
  
\bibitem{Hellerman:2002ax}
  S.~Hellerman, J.~McGreevy and B.~Williams,
  JHEP {\bf 0401} (2004) 024
  [hep-th/0208174].
  
\bibitem{Hellerman:2006tx}
  S.~Hellerman and J.~Walcher,
  ``Worldsheet CFTs for Flat Monodrofolds,'' \\  
  hep-th/0604191.
   
\bibitem{Schulgin:2008fv}
  W.~Schulgin and J.~Troost,
  JHEP {\bf 0812} (2008) 098
  [arXiv:0808.1345 [hep-th]].
  
\bibitem{Condeescu:2012sp}
  C.~Condeescu, I.~Florakis and D.~L\"ust,
  JHEP {\bf 1204} (2012) 121 \\  
{}  [arXiv:1202.6366 [hep-th]].
  
\bibitem{Condeescu:2013yma}
  C.~Condeescu, I.~Florakis, C.~Kounnas and D.~L\"{u}st,
  JHEP {\bf 1310} (2013) 057 \\ {}
  [arXiv:1307.0999 [hep-th]].
 
\bibitem{McOrist:2010jw}
  J.~McOrist, D.~R.~Morrison and S.~Sethi,
  Adv.\ Theor.\ Math.\ Phys.\  {\bf 14} (2010)\\ {}
  [arXiv:1004.5447 [hep-th]].
  
\bibitem{deBoer:2012ma}
  J.~de Boer and M.~Shigemori,
  Phys.\ Rept.\  {\bf 532} (2013) 65
  [arXiv:1209.6056 [hep-th]].
  
\bibitem{Dibitetto:2012rk}
  G.~Dibitetto, J.~J.~Fernandez-Melgarejo, D.~Marques and D.~Roest, \\
  Fortsch.\ Phys.\  {\bf 60} (2012) 1123
  [arXiv:1203.6562 [hep-th]].
  
  \bibitem{wein}
Z.-J.~Liu, A.~Weinstein and P.~Xu, 
J. Diff. Geom. {\bf 45} (1997) 547 [dg-ga/9508013];\\
\dittostraight,~
Commun. Math. Phys. {\bf 192} (1998) 121 [dg-ga/9611001].

  \bibitem{dirac}
 T.~Courant,
 Trans. Amer. Math. Soc. {\bf 319} (1990) 631.

 \bibitem{Roy}
D.~Roytenberg, ``Courant algebroids, derived brackets and even symplectic supermanifolds,'' Ph.D. thesis, math/9910078 [math.DG];\\
\dittostraight,~ Lett. Math. Phys. {\bf 61} (2002) 123 [math/0112152 [math.QA]]

\bibitem{Roytenberg:2006qz}
  D.~Roytenberg,
  Lett.\ Math.\ Phys.\  {\bf 79} (2007) 143
  [hep-th/0608150].

\bibitem{Alexandrov:1995kv}
  M.~Alexandrov, M.~Kontsevich, A.~Schwartz and O.~Zaboronsky, \\
    Int.\ J.\ Mod.\ Phys.\ A {\bf 12} (1997) 1405
  [hep-th/9502010].

\bibitem{ikeda}
  N.~Ikeda,
  ``Lectures on AKSZ Topological Field Theories for Physicists,''\\ {}
  arXiv:1204.3714 [hep-th].

 \bibitem{severa}
H.~Bursztyn, M.~Crainic and P.~\v{S}evera, 
Travaux math\'ematiques {\bf 16} (2005) 41.

\bibitem{sw}
 P.~\v{S}evera and A.~Weinstein,
 Prog. Theor. Phys. Suppl. {\bf 144} (2001) 145 \\ {}[math/0107133 [math.SG]]. 
  
\bibitem{Hofman:2002rv}
  C.~Hofman and J.~S.~Park,
  ``Topological open membranes,''
  hep-th/0209148.
  
  
\bibitem{Tseytlin:1990nb}
  A.~A.~Tseytlin,
  Phys.\ Lett.\ B {\bf 242} (1990) 163;\\
  \dittostraight,~
  Nucl.\ Phys.\ B {\bf 350} (1991) 395.
  
\bibitem{Siegel:1993xq}
  W.~Siegel,
  Phys.\ Rev.\ D {\bf 47} (1993) 5453  [hep-th/9302036]; \\
 \dittostraight,~
  Phys.\ Rev.\ D {\bf 48} (1993) 2826
  [hep-th/9305073];\\
  \dittostraight,~
  In *Berkeley 1993, Proceedings, Strings '93* 353-363, and State U. New York Stony Brook - ITP-SB-93-050 (93,rec.Sep.) 11 p. (315661)
  [hep-th/9308133].

    
\bibitem{Hull:2009mi}
  C.~Hull and B.~Zwiebach,
  JHEP {\bf 0909} (2009) 099
  [arXiv:0904.4664 [hep-th]].

\bibitem{Hohm:2010jy}
  O.~Hohm, C.~Hull and B.~Zwiebach,
  JHEP {\bf 1007} (2010) 016
  [arXiv:1003.5027 [hep-th]].

\bibitem{Hohm:2010pp}
  O.~Hohm, C.~Hull and B.~Zwiebach,
  JHEP {\bf 1008} (2010) 008
  [arXiv:1006.4823 [hep-th]].
  
\bibitem{Aldazabal:2013sca}
  G.~Aldazabal, D.~Marques and C.~Nu\~nez,
  Class.\ Quant.\ Grav.\  {\bf 30} (2013) 163001\\ {}
  [arXiv:1305.1907 [hep-th]].

\bibitem{Berman:2013eva}
  D.~S.~Berman and D.~C.~Thompson,
  Phys.\ Rept.\  {\bf 566} (2014) 1 \\  
  {} [arXiv:1306.2643 [hep-th]].

\bibitem{Hohm:2013bwa}
  O.~Hohm, D.~L\"ust and B.~Zwiebach,
  Fortsch.\ Phys.\  {\bf 61} (2013) 926 \\  
{}  [arXiv:1309.2977 [hep-th]].
  
\bibitem{dftcft}
  R.~Blumenhagen, F.~Hassler and D.~L\"ust,
  JHEP {\bf 1502} (2015) 001 \\  
{}  [arXiv:1410.6374 [hep-th]].
  
\bibitem{dftcft2}
 R.~Blumenhagen, P.~d.~Bosque, F.~Hassler and D.~Lust,
  JHEP {\bf 1508} (2015) 056
  [arXiv:1502.02428 [hep-th]].
  
\bibitem{Berman:2007xn}
  D.~S.~Berman, N.~B.~Copland and D.~C.~Thompson,
  Nucl.\ Phys.\ B {\bf 791} (2008) 175\\ {}
  [arXiv:0708.2267 [hep-th]].
  
\bibitem{Dall'Agata:2008qz}
  G.~Dall'Agata and N.~Prezas,
  JHEP {\bf 0808} (2008) 088
  [arXiv:0806.2003 [hep-th]].
  
\bibitem{Avramis:2009xi}
  S.~D.~Avramis, J.~P.~Derendinger and N.~Prezas,
  Nucl.\ Phys.\ B {\bf 827} (2010) 281
  [arXiv:0910.0431 [hep-th]].

  
\bibitem{dftflux1}
  G.~Aldazabal, W.~Baron, D.~Marques and C.~Nu\~nez,
  JHEP {\bf 1111} (2011) 052 \\  
{}   [Erratum-ibid.\  {\bf 1111} (2011) 109]
  [arXiv:1109.0290 [hep-th]].
  
\bibitem{dftflux2}
  D.~Geissbuhler,
  JHEP {\bf 1111} (2011) 116
  [arXiv:1109.4280 [hep-th]].
  
\bibitem{dftflux3}
  D.~Geissbuhler, D.~Marques, C.~Nu\~nez and V.~Penas,
  JHEP {\bf 1306} (2013) 101\\ {}
  [arXiv:1304.1472 [hep-th]].
  
\bibitem{dftflux4}
  R.~Blumenhagen, X.~Gao, D.~Herschmann and P.~Shukla,
  JHEP {\bf 1310} (2013) 201\\ {}
  [arXiv:1306.2761 [hep-th]].

\bibitem{fluxa}
  D.~Andriot, M.~Larfors, D.~L\"ust and P.~Patalong,
  JHEP {\bf 1109} (2011) 134\\ {}
  [arXiv:1106.4015 [hep-th]].
  
\bibitem{fluxb}
  D.~Andriot, O.~Hohm, M.~Larfors, D.~L\"ust and P.~Patalong,\\
    Fortsch.\ Phys.\  {\bf 60} (2012) 1150
  [arXiv:1204.1979 [hep-th]].
  
\bibitem{fluxd}
  D.~S.~Berman, E.~T.~Musaev and D.~C.~Thompson, \\
    JHEP {\bf 1210} (2012) 174
  [arXiv:1208.0020 [hep-th]].
  
\bibitem{fluxc}
  D.~Andriot and A.~Betz,
  JHEP {\bf 1312} (2013) 083
  [arXiv:1306.4381 [hep-th]].
  
\bibitem{Deser:2014mxa}
 A.~Deser and J.~Stasheff,
  Commun.\ Math.\ Phys.\  {\bf 339} (2015) 3\\ {}
  [arXiv:1406.3601 [math-ph]].
  
\bibitem{KS}
Y.~Kosmann-Schwarzbach,
in {\it{The Breadth of Symplectic and Poisson Geometry}},\\
Progr. Math. {\bf 232} (2005) 363
[arXiv:math/0310359].


 \bibitem{MckenzieXu}
 K.~C.~H.~Mackenzie and P.~Xu,
 Duke Math. J. {\bf 73} (1994) 415. 

\bibitem{Chatzistavrakidis:2014tda}
  A.~Chatzistavrakidis,
  Phys.\ Rev.\ D {\bf 90} (2014)  024038
  [arXiv:1404.2812 [hep-th]].
 
 \bibitem{Rieffel}
M.~A.~Rieffel, 
Commun. Math. Phys. {\bf 122}  (1989)  531.  
 
\bibitem{Cattaneo:2009zx}
  A.~S.~Cattaneo, J.~Qiu and M.~Zabzine,
  Adv.\ Theor.\ Math.\ Phys.\  {\bf 14} (2010) 695\\ {}
  [arXiv:0911.0993 [hep-th]].

 
\bibitem{Kotov:2004wz}
  A.~Kotov, P.~Schaller and T.~Strobl,
  Commun.\ Math.\ Phys.\  {\bf 260} (2005) 455 \\  
{}  [hep-th/0411112].
  
\bibitem{Salnikov:2013pwa}
  V.~Salnikov and T.~Strobl,
  JHEP {\bf 1311} (2013) 110
  [arXiv:1311.7116 [math-ph]].

  \bibitem{psm}
 C.~Klim{\v c}\'ik and T.~Strobl,
 J. Geom. Phys. \textbf{43} (2002) 341 \\ 
{} [math/0104189 [math.SG]].


\bibitem{CAred1}
  P.~Bouwknegt, J.~Evslin and V.~Mathai,
  Commun.\ Math.\ Phys.\  {\bf 249} (2004) 383\\ {}
  [hep-th/0306062].
  
\bibitem{CAred2}
  P.~Bouwknegt, J.~Evslin and V.~Mathai,
  Phys.\ Rev.\ Lett.\  {\bf 92} (2004) 181601
 \newline [hep-th/0312052].
  
\bibitem{CAred3}
  H.~Bursztyn, G.~R.~Cavalcanti and M.~Gualtieri,
  Adv.\ Math.\  {\bf 211} (2007) 726
 \newline [math/0509640 [math.DG]].
  
\bibitem{CAred4}
  G.~R.~Cavalcanti and M.~Gualtieri,
  A Celebration of the Mathematical Legacy of Raoul Bott (CRM Proceedings \& Lecture Notes) American Mathematical Society (2010) 341-366.
  [arXiv:1106.1747 [math.DG]].
  
\bibitem{CAred5}
  P.~\v{S}evera,
Lett. Math. Phys. {\bf 105} (2015) 1689 [arXiv:1502.04517 [math.SG]].


\bibitem{sixmanifolds}
G.~R.~Cavalcanti and M.~Gualtieri,
J. Symplectic Geom. \textbf{2} (2004)  393 \\
{} [math/0404451 [math.DG]].

 \end{thebibliography}
\end{document}